# Cross-frequency interactions during diffusion on complex brain networks are facilitated by scale-free properties

Roberto C. Sotero[1*], Lazaro M. Sanchez-Rodriguez[1], Mehdy Dousty[2], Yasser Iturria-Medina[3], Jose M. Sanchez-Bornot[4], for the Alzheimer's Disease Neuroimaging Initiative[§]

[1]Hotchkiss Brain Institute, and Department of Radiology, University of Calgary, Calgary, Canada, T2N 4N1

[2]Institute of Biomaterials and Biomedical Engineering, University of Toronto, Toronto, ON, Canada

[3]Montreal Neurological Institute, McGill University, Montreal, Canada, H3A 2B4

[4]Intelligent Systems Research Centre, Ulster University, Magee Campus, Derry Londonderry, UK

[*]Correspondence: roberto.soterodiaz@ucalgary.ca

[§]Data used in preparation of this article were obtained from the Alzheimer's Disease Neuroimaging Initiative (ADNI) database (adni.loni.usc.edu). As such, the investigators within the ADNI contributed to the design and implementation of ADNI and/or provided data but did not participate in analysis or writing of this report. A complete listing of ADNI investigators can be found at: http://adni.loni.usc.edu/wp-content/uploads/how_to_apply/ADNI_Acknowledgement_List.pdf.

**Abstract**

We studied the interactions between different temporal scales of diffusion processes on complex networks and found them to be stronger in scale-free (SF) than in Erdos-Renyi (ER) networks, especially for the case of phase-amplitude coupling (PAC)— the phenomenon where the phase of an oscillatory mode modulates the amplitude of another oscillation. We found that SF networks facilitate PAC between slow and fast frequency components of the diffusion process, whereas ER networks enable PAC between slow-frequency components. Nodes contributing the most to the generation of PAC in SF networks were non-hubs that connected with high probability to hubs. Additionally, brain networks from healthy controls (HC) and Alzheimer's disease (AD) patients presented a weaker PAC between slow and fast frequencies than SF, but higher than ER. We found that PAC decreased in AD compared to HC and was more strongly correlated to the scores of two different cognitive tests than what the strength of functional connectivity was, suggesting a link between cognitive impairment and multi-scale information flow in the brain.

## Introduction

The study of information flow and transport in complex biological and social networks by means of diffusion processes has attracted increasing interest in recent years[1–4]. Random walks[5], the processes by which randomly-moving objects wander away from their starting location are commonly used to describe diffusion processes. In the past decades, there has been considerable progress in characterizing first passage times, or the amount of time it takes a random walker to reach a target [6–9]. However, previous works have neglected the study of the temporal dynamics of the information flow in the network, which depends on how the walkers move and not just on their arrival time. Thus, we lack knowledge about how the different temporal scales in the diffusion processes arise from the topological structure of the network, whether they interact, and how they do it.

Diffusion processes on different complex networks may have associated multiple temporal scales. Popular methods such as Fourier transform allow for the identification of main oscillatory features of a system but struggle with nonlinear and nonstationary signals[10]. These issues can be addressed by using empirical mode decomposition (EMD)[10], an adaptive and data-driven method that decomposes nonlinear and nonstationary signals, like the movement of the random walkers, into fundamental modes of oscillations called intrinsic mode functions (IMFs), without the need for a predefined model as is the case for Fourier and wavelet transforms.

In this paper we study the interaction between IMFs extracted from diffusion processes occurring in simulated Erdos-Renyi (ER) [11] and scale-free (SF)[12] networks, as well as in real brain networks estimated from resting-state functional magnetic resonance imaging (rs-fMRI) and diffusion weighted magnetic resonance imaging (DWMRI) data recorded from healthy subjects and patients with Alzheimer's disease (AD). Since IMFs are associated with different oscillatory modes, their interactions correspond to the phenomenon known as cross-frequency coupling (CFC)[13]. We focus on three types of CFC: phase-amplitude coupling (PAC), the phenomenon where the instantaneous phase of a low frequency oscillation modulates the instantaneous amplitude of a higher frequency oscillation [14 15]; amplitude-amplitude coupling (AAC), which measures the co-modulation of the instantaneous amplitudes of two oscillations[16]; and phase-phase coupling (PPC), which corresponds to the synchronization between two instantaneous phases[17].

## Results

### Diffusion of simulated ER and SF networks

We start by considering an unweighted network consisting of $N$ nodes. We place a large number $K (K \gg N)$ of random walkers onto this network. At each time step, the walkers move randomly between the nodes that are directly linked to each other. We allow the walkers to perform $T$ time steps. As a walker visits a node, we record the fraction of walkers present at it, which we term node activity. Thus, after $T$ time steps, we obtain $K$ time series reflecting different realizations of the flow of information in the network.

Two types of simulated complex networks are considered here, ER and SF networks. An ER network is a random graph where each possible edge has the same probability $p$ of existing. The degree of a node $i$ ($k_i$) is defined as the number of connections it has to other nodes. The degree distribution $P(k)$ of an ER network is a binomial distribution, which decays exponentially for large degrees $k$, allowing only very small degree fluctuations[18]. On the other hand, SF networks are constructed with the Barabasi and Albert's (BA) model[12], or "rich-gets-richer" scheme, which assumes that new nodes in a network are not connected at random but with high probability to those which already possess a large number of connections (also known as hubs). In the BA model, $P(k)$ decays as a power law, which yields scale-invariance and allows for large degree fluctuations. We generate ER and SF networks by means of the MATLAB (The MathWorks Inc., Natick, MA, USA) toolbox CONTEST[19].

Figures 1a and 1c show an example of connectivity (adjacency) matrices for ER and SF networks, respectively. Both networks have the same number of edges $m$, and nodes, corresponding to a sparsity, $e$, value of $e = 1 - \frac{m}{N^2} = 0.9$. A number of $10^4$ random walkers were placed onto these networks and diffused for 5000 time steps. One realization of node activity is shown in Fig. 1b and Fig. 1d for ER and SF networks, respectively. We then applied a recent version of the EMD method [20] to these two time series (see *Materials and Methods*). Figure 1e shows the first 7 IMFs and residue $(R)$ for the ER and SF networks. The first IMF (IMF1) corresponds to the fastest oscillatory mode and the last IMF to the slowest one. Note that IMF7 is the sum of all the slow IMFs up to IMF7. As seen in Fig. 1e, the EMD method produces amplitude and frequency modulated signals. By applying the Hilbert transform to each IMF, instantaneous amplitudes, phases, and frequencies can be obtained and a time-frequency representation of the original signal (known as the Hilbert spectrum) can be constructed[10]. Since each time instant in Fig. 1e corresponds to a different node in the network, computing the instantaneous frequencies of the 7

IMFs yields a Hilbert spectrum [10] that links the frequency of the oscillatory modes recorded at each node to the node's degree. Fig. 1f and Fig 1g show the Hilbert spectrum for the ER (Fig 1a) and SF (Fig 1c) networks, respectively. The color scale represents the energy of the spectrum. Our results show that ER networks have more energy in the low frequencies and present a narrow range of node degrees. On the other hand, SF networks present a wide distribution of node degree values where nodes with low degrees are more associated to low frequency oscillations, whereas high degree nodes relate to high frequencies. These results indicate that random walkers strongly link low and high frequency dynamics when they diffuse in SF networks.

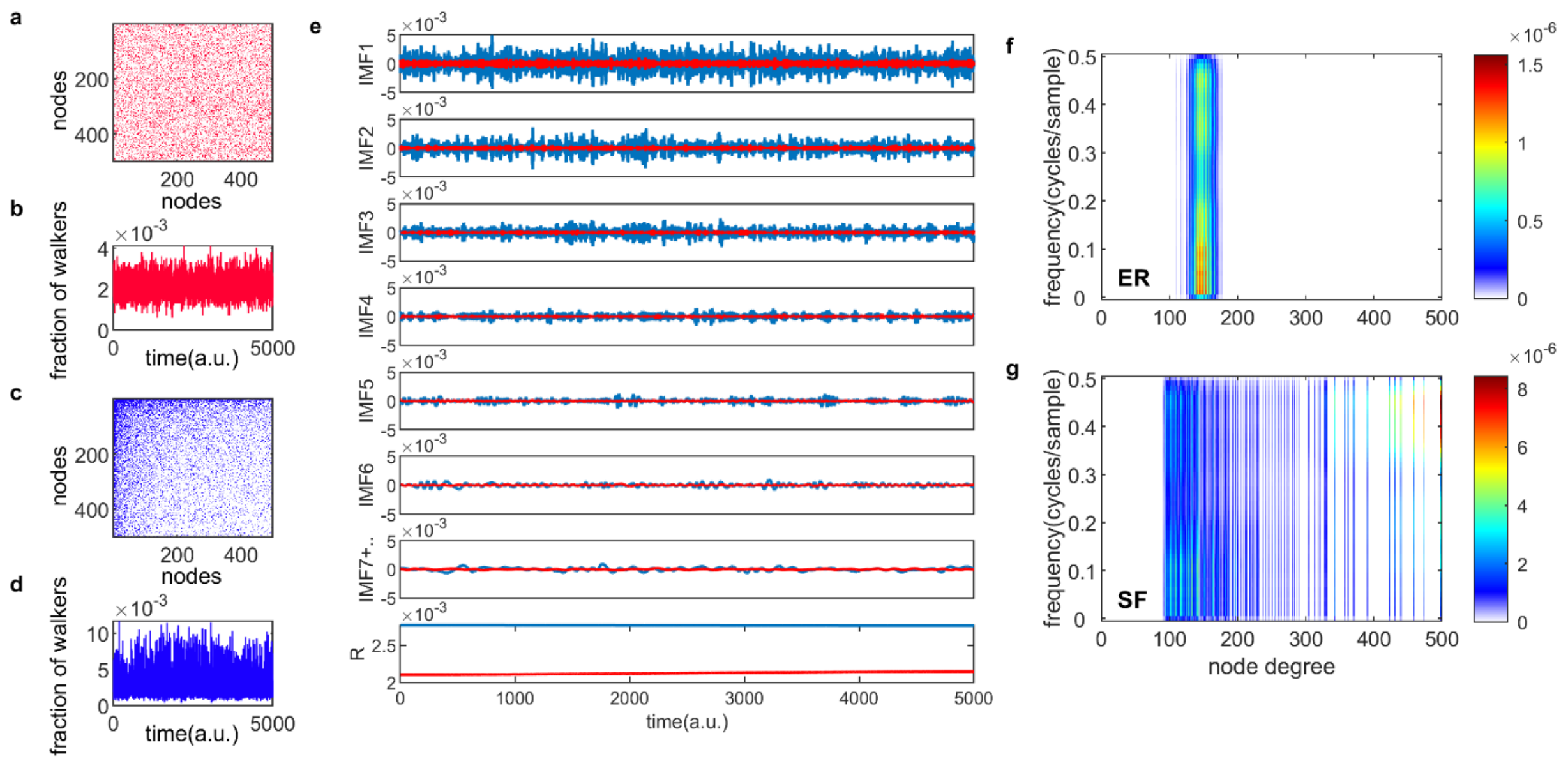

**Fig. 1.** Different temporal modes of diffusion on complex networks. Panels **a** and **c** are examples of ER and SF connectivity matrices, respectively, with $N = 500$ and $e = 0.9$. Panels **b** and **d** show one realization of node activity, i.e., the fraction of the rest of the walkers seen by one walker at each node it lands at during in the network for 5000 time steps, for ER and SF networks, respectively. The total number of walkers present in each network was $10^4$. Panel **e** shows the empirical mode decomposition (EMD) of the time series in **b** and **d**, producing different intrinsic mode functions (IMF) and a residue (R). This appears in red (blue) for the ER (SF) networks. Panels **f**, and **g** show the spectrum of the diffusion process organized by the node degree for ER and SF networks, respectively.

To characterize the interaction between frequencies, we computed three types of CFC interactions, $PAC_{kl}$, $AAC_{kl}$, and $PPC_{kl}$, between all possible combinations of the 7 IMFs ($k = 1,2,\dots,7$, $l = 1,2,\dots,7$, $k > l$, thus obtaining a 7x7 upper triangular matrix for each measure) for 3 different values of sparsity ($e = [0.9,0.8,0.7]$) of ER and SF networks (Fig. 2). Figure 2 shows the average over $K = 10^4$ realizations of $PAC_{kl}$ (Fig. 2a), $AAC_{kl}$ (Fig. 2b), and $PPC_{kl}$ (Fig. 2c) for ER and SF networks. Figure S1 shows the corresponding $Z$-scores. In the case of $PAC$, $Z$-score values obtained for SF networks were higher than the corresponding values obtained in ER networks. Strong $PAC$ values in SF networks involved the phase of IMF7 (the slowest IMF) and the amplitudes of IMF6 to IMF1. On the other hand, the highest $AAC$ and $PPC$ values in SF networks involved IMFs with close frequencies such as IMF1 and IMF2. In the case of ER networks, the strongest values were obtained for interactions between slow IMFs for $PAC$ (phase of IMF7 and amplitude of IMF5 in Fig. 2a), $AAC$ (amplitudes of IMF7 and IMF6 in Fig. 2b) and between fast IMFs for $PPC$ (phases of IMF2 and IMF1 in Fig. 2c). When we decreased the level of sparsity (i.e. the network became more connected), the results for SF networks turned similar to the ones in ER networks. In conclusion, $PAC$ interactions in SF networks were the strongest CFC found (as reflected by the $Z$-scores) and, when compared to results from ER networks, the main difference was the existence of strong $PAC$ between slow and fast oscillatory components of the diffusion process.

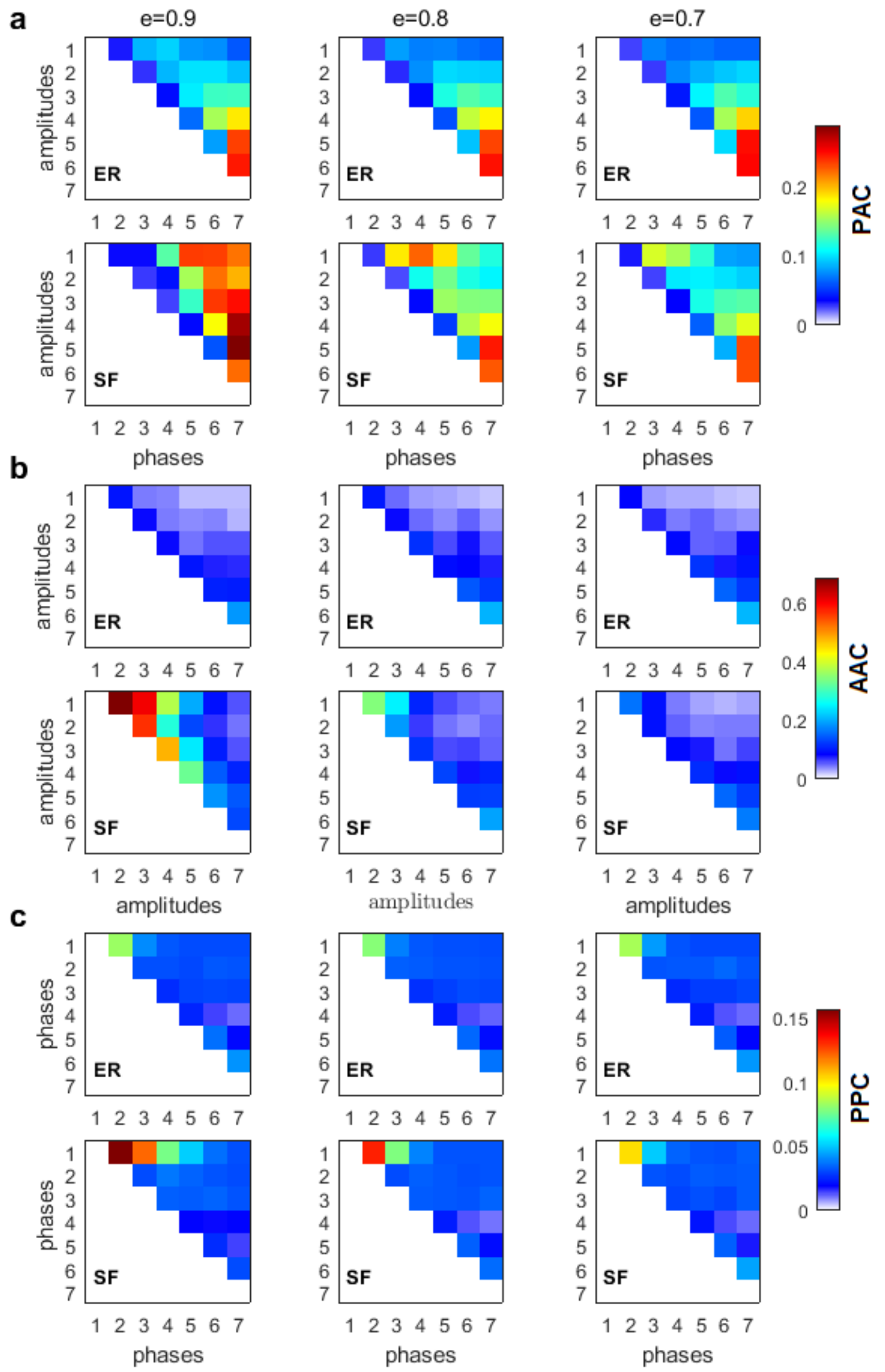

**Fig. 2.** Cross-frequency interactions between the fundamental modes of diffusion on ER and SF networks. All simulated networks had $N = 500$ nodes, and $10^4$ random walkers were placed over them, each performing $5000$ time steps. Three different values of network sparsity were considered: $e = [0.9,0.8,0.7]$. (**a**) phase-amplitude coupling (PAC), (**b**) amplitude-amplitude coupling (AAC), (**c**) phase-phase coupling (PPC).

To verify our results were not an artifact of the application of the EMD method, we defined seven non-overlapping frequency bands (0.001-0.009, 0.010-0.020, 0.021-0.040, 0.041-0.060, 0.061-0.100, 0.101-0.250, and 0.251-0.490 cycles/sample) based on the seven IMFs and computed the CFC measures. Figure S2 shows similar CFC patterns to the ones obtained for the Z-scores in figure S1, suggesting our results are not dependent on the EMD method but a consequence of the network architecture instead. Differences between the two figures are associated to the fact that consecutive IMFs have a small overlap in frequency by design[10].

We also studied the influence of specific nodes in the SF networks in the generation of PAC, AAC and PPC. The contribution of each node $i$ was computed by removing the node from the network and running the random walker analysis on the new network. The obtained PAC, AAC and PPC were denoted as $PAC_i^r$, $AAC_i^r$, and $PPC_i^r$, respectively. The contribution of a node to the corresponding CFC measure is the change in the CFC value as a result of removing the node from the network: $\Delta PAC_i = |PAC_i^r/PAC - 1|$, $\Delta AAC_i = |AAC_i^r/AAC - 1|$, and $\Delta PPC = |PPC_i^r/PPC - 1|$. Additionally, we computed several local topological properties for all nodes in the network using the Brain Connectivity Toolbox[21], namely: the degree $(k)$; the efficiency $(e)$, which quantifies a network's resistance to failure on a small scale; the clustering coefficient $(cc)$, which measures the degree to which nodes in a graph tend to cluster together; assortativity $(a)$, which indicates if a node tends to link to other nodes with the same or similar degree; betweenness centrality $(bc)$, which is the fraction of shortest paths in the network that contain a given node (a node with higher betweenness centrality has more control over the network because more information will pass through it); eigenvector centrality $(ec)$, which is another measure of centrality where relative scores are assigned to all nodes based on the concept that connections to high-scoring nodes contribute more to the score of the node in question than equal connections to low-scoring nodes; subgraph centrality $(sc)$, which is a weighted sum of closed walks of different lengths in the network starting and ending at the node; and the product of the three centrality measures $(ec * sc * bc)$. Figure 3a, 3b, and 3c show the Pearson correlation between the eight topological measures and ΔPAC, ΔAAC and ΔPPC, respectively. Different frequency combinations presented different correlation values. The strongest correlations involving ΔPAC were found for the topological measure composed by the product of the three centrality measures $(ec * sc * bc)$, between the phase of $IMF3$ and the amplitude of $IMF1$. The amplitude of $IMF1$ was also involved in strong correlations with the phases of $IMF4$, $IMF5$, and $IMF6$. Of the three CFC measures, ΔAAC was most strongly correlated to topology (Fig. 3b), specifically with centrality measures, followed by ΔPAC. ΔPPC was weakly correlated to the topology of the network.

Next, by using the degrees $k$ we classified nodes in the network into hubs if their degree was at least one standard deviation above the network mean[22], and into non-hubs otherwise. We then computed the average ΔPAC of all frequency combinations involving the amplitudes of fast frequencies ($IMF1$ and $IMF2$) and the phases of slow frequencies ($IMF5$, $IMF6$, and $IMF7$). Note that the correlations between ΔPAC corresponding to these frequency combinations and the product $ec * sc * bc$ are between 0.24 and 0.47 (Fig. 3a), which suggests that other mechanisms are needed to explain these ΔPAC values.

Figure 4a plots ΔPAC versus node degree for all nodes in the SF network. Interestingly, hubs, the most connected nodes in the network, are not necessarily involved in the largest ΔPAC values. Non-hubs were classified into three groups by equally dividing the ΔPAC range $(0 - 0.6)$: bottom $(0 - 0.2)$, middle $(0.2 - 0.4)$, and top $(0.4 - 0.6)$. The histogram in Fig. 4b shows the probability that nodes in the four groups (hubs and three non-hubs groups) have of connecting to nodes of certain degrees. We see that top non-hubs connect to high degree nodes (hubs in Fig. 4a) with higher probability than middle and bottom non-hubs. On the other hand, hubs connect with high probability to low degree nodes. Since PAC is defined as the coupling from a low to a high frequency, its highest contributor will be the nodes associated more with low frequencies (i.e. nodes with low degrees, see Fig. 1g) and that also connect to nodes that are more associated to high frequencies (i.e. nodes with high degrees, see Fig 1g); that is, the top non-hubs. Accordingly, hubs, which are more connected to low frequency nodes contribute less to PAC, except for only one hub which presented the largest $\Delta PAC_i$ of all nodes in the network (Fig. 1a, c, and e). This hub (node with degree 270 in Fig. 4) is known as a super-hub for having degree significantly higher than other hubs in the network [23]. Since the classification into top, middle and bottom non-hubs based on $\Delta\mathrm{PAC}$ values is somewhat arbitrary, we explored the results of changing the $\Delta\mathrm{PAC}$ range of these three groups. Figures 4c and 4d show the results when the groups were defined by the bands: bottom $(0-0.1)$, middle $(0.1-0.5)$, and top $(0.5-0.6)$. In this case, the number of nodes in the top and bottom groups were reduced and the probability that top non-hubs connected to high degree nodes increased.

In the calculations leading to Fig. 4e we increased the number of nodes in the top and bottom groups as compared to Fig. 4a by selecting the ranges: bottom $(0-0.35)$, middle $(0.35-0.45)$, and top $(0.45-0.6)$. In this case, the probability for the top non-hubs decreased and the results for the top and middle groups were more similar (see Fig. 4f).

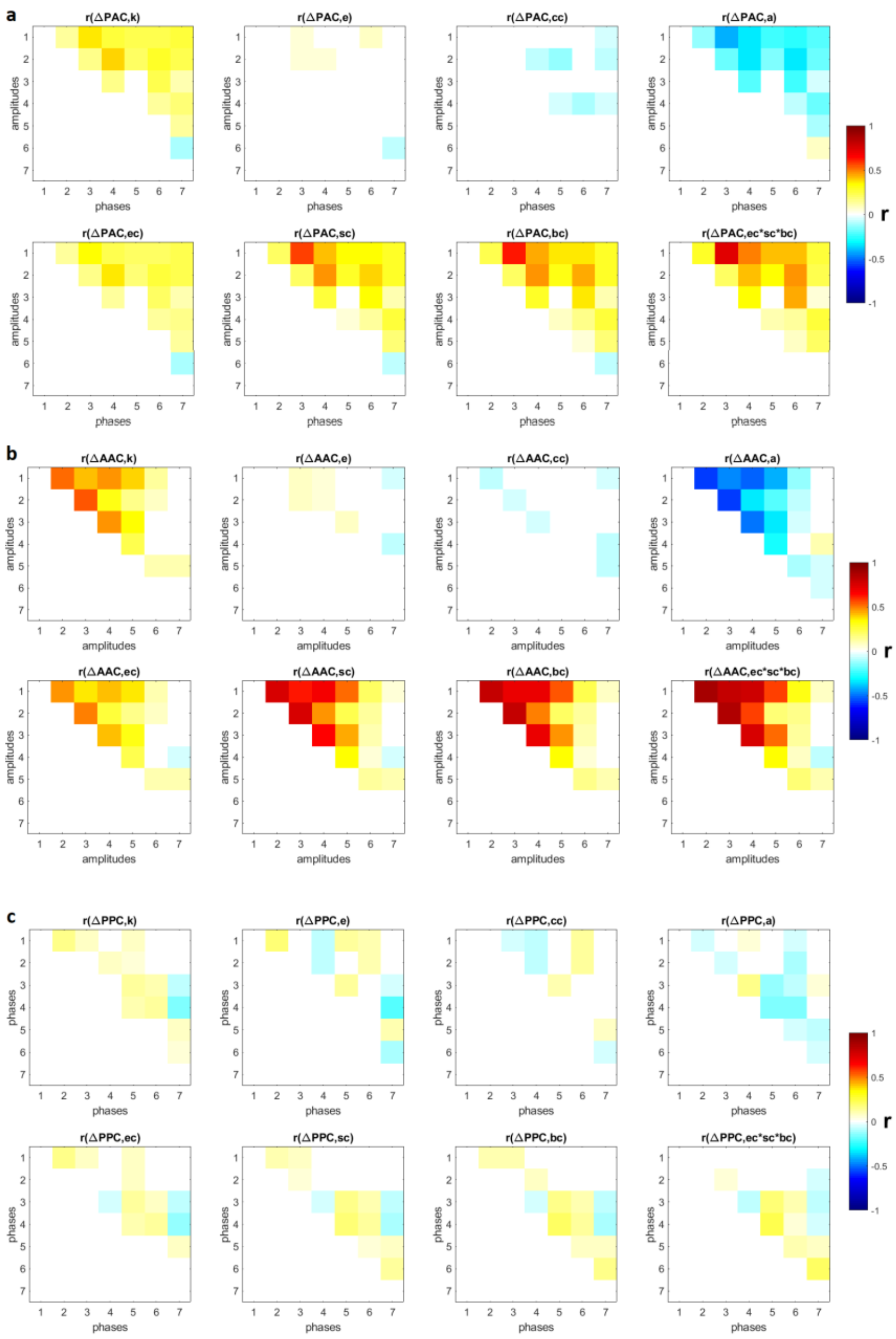

**Fig. 3.** Correlation between changes in CFC and eight topological measures: degree $(k)$, efficiency $(e)$, clustering coefficient $(cc)$, assortativity $(a)$, eigenvector centrality $(ec)$, subgraph centrality $(sc)$, betweenness centrality $(bc)$, product of three centrality measures $(ec * sc * bc)$. Non-significant $(p < 0.05)$ correlation values after correction by false-discovery rate are displayed in white. (**a**) PAC, (**b**) AAC, (**c**) PPC.

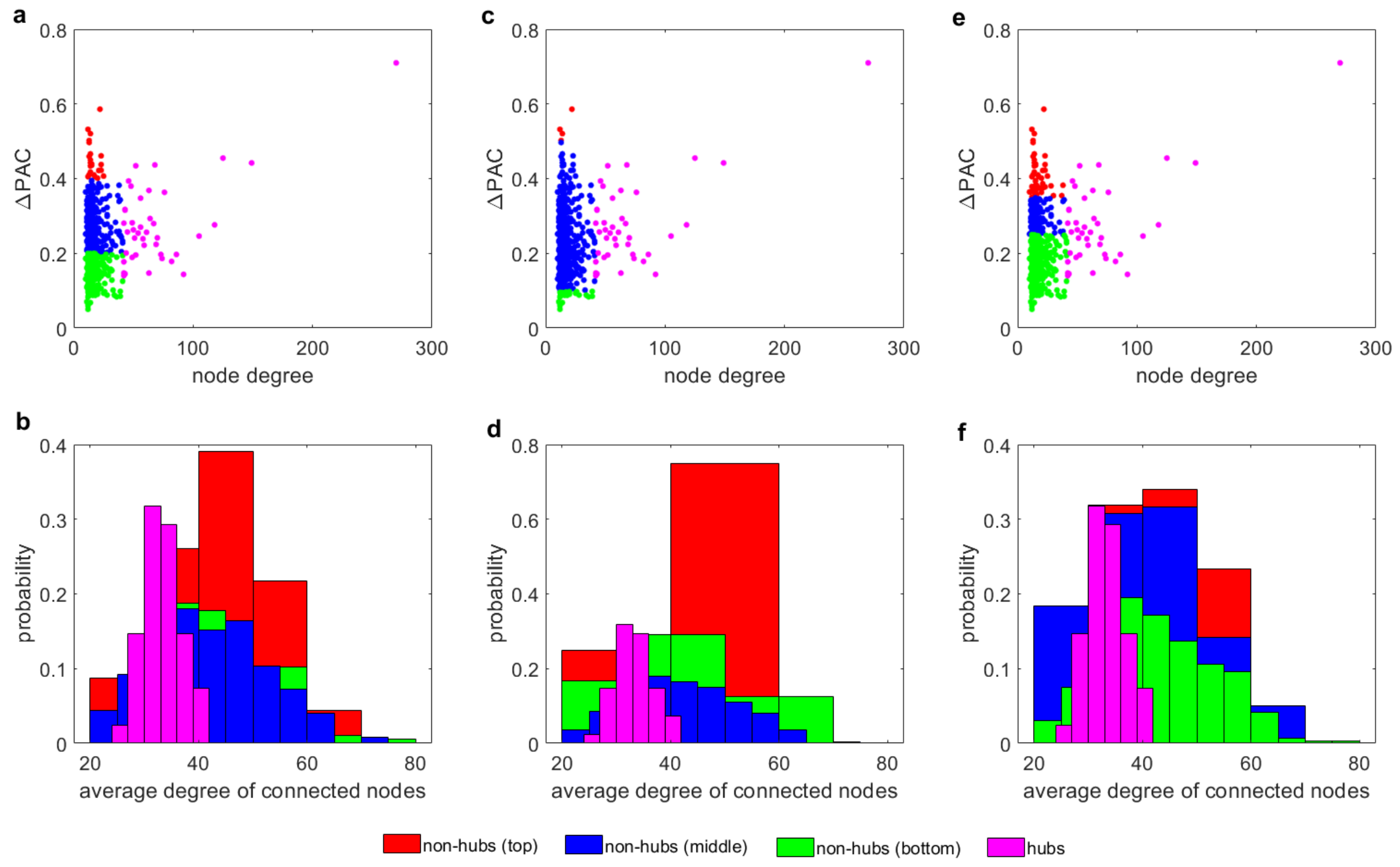

**Fig. 4.** The influence of non-hubs vs hubs on ΔPAC. Average ΔPAC for each node degree for three different grouping of non-hubs: (**a**) bottom $(0-0.2)$, middle $(0.2-0.4)$, and top $(0.4-0.6)$, (**b**) bottom $(0-0.1)$, middle $(0.1-0.5)$, and top $(0.5-0.6)$, and (**c**) bottom $(0-0.25)$, middle $(0.25-0.35)$, and top $(0.35-0.6)$. Panels **b**, **d**, and **f** present the probability of the four different groups of nodes of connecting to nodes of certain degrees, corresponding to the node distribution presented in **a**, **c**, and **d**, respectively.

## Diffusion on brain networks estimated from healthy and Alzheimer's disease subject's data

The information flow, as given by the movement of the random walkers, was also investigated in real brain networks. For this, freely available (http://adni.loni.usc.edu) images from the Alzheimer's disease neuroimaging initiative (ADNI) were utilized. In a first stage, structural magnetic resonance images (MRI) and DWMRI obtained for 51 healthy control (HC) subjects were used to estimate anatomical connectivity matrices for each subject [24,25] (see Materials and Methods). A network backbone containing the dominant connections in the average network was computed using a minimum-spanning tree based algorithm [26]. The anatomical backbone was then transformed into a matrix of zeros (no connection existing between two nodes) and ones (a link exists). Moreover, Blood Oxygen Level Dependent (BOLD) signals were obtained from rs-fMRI (see Methods) for 31 Alzheimer's disease (AD) patients and 44 HCs from ADNI and used to compute functional connectivity (FC) matrices by taking the absolute value of the Pearson correlation. Each FC matrix was multiplied by the anatomical backbone, resulting in a new matrix we denote as $\boldsymbol{W}$. Thus, the random walkers flow in the structural network, but their movement is influenced by the brain's activity. This guarantees that the dynamics of the information flow will change if instead of the resting-state we study a different condition such as stimulation or anesthesia [27,28].

Figures 5a and 5e show the connectivity matrix $\boldsymbol{W}$ for a representative HC and an AD patient, respectively. For each subject, we placed $10^4$ random walkers on top of its $\boldsymbol{W}$ and let them diffuse during $5000$ time steps. The transition probability $p_{ij}$ from brain area $i$ to brain area $j$ is given by $p_{ij} = \frac{w_{ij}}{\sum_{j=1}^{N} w_{ij}}$, where $w_{ij}$ is the weight of the connection from area $i$ to area $j$ [29]. Since both the FC and the anatomical backbone are symmetric matrices, we have: $w_{ij} = w_{ji}$, and $p_{ij} = p_{ji}$.

For each AD and HC subject, we obtained 8 IMFs from the time series generated by the activity of the random walkers. We then focused on PAC since it was the strongest CFC type obtained for both ER and SF networks in our simulations (Supplementary Fig. 1). Figure 5b shows the PAC between all possible combinations of the 8 IMFs, averaged over $10^4$ realizations and over the 44 HC subjects, denoted as $PAC_{HC}$. The strongest PAC values were obtained for interactions between slow IMFs (the phase of IMF8 and the amplitudes of IMF5, IMF6 and IMF7). We also computed the average PAC across AD patients $(PAC_{AD})$ and compared it to the HC group by computing the measure $\frac{PAC_{AD}}{PAC_{HC}} - 1$. Our results (Fig. 5f) show that PAC between fast frequencies (IMF1) and slower modes (IMFs 3 to 8) weaken during AD as compared to HCs.

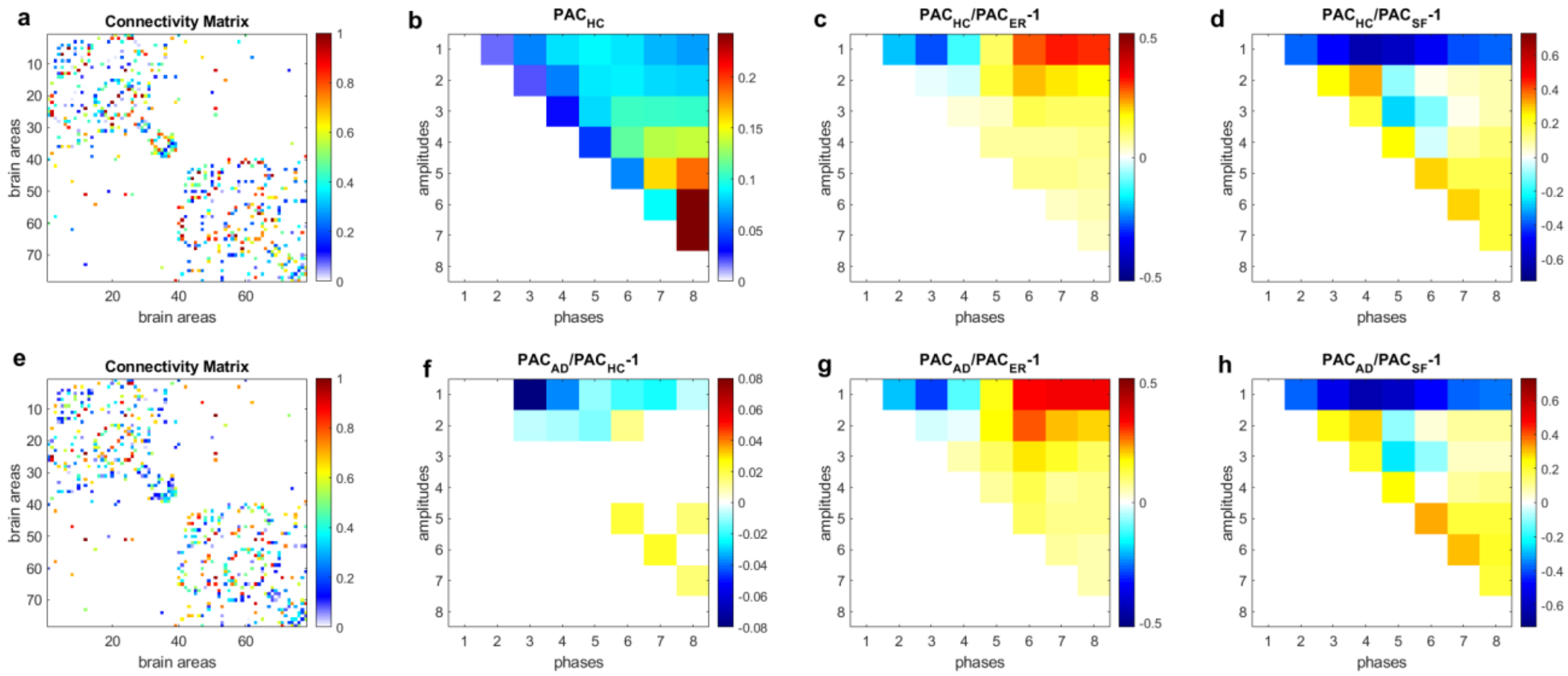

**Fig. 5.** PAC in HC and AD. Connectivity matrix of a HC (**a**) and AD (**e**) subject. PAC averaged over HC subjects (**b**) and comparison with AD subjects (**f**). Comparison of PAC from HC and PAC generated from equivalent ER (**c**) and SF (**d**) networks. Comparison of PAC from AD and PAC generated from equivalent ER (**g**) and SF (**h**) networks.

Additionally, for each subject, we generated 500 ER and 500 SF networks of the same size and number of edges as their $\boldsymbol{W}$ matrices, computed PAC for these matrices and averaged the results, obtaining $PAC_{ER}$, and $PAC_{SF}$, respectively. We then computed the following measures: $\frac{PAC_{HC}}{PAC_{ER}} - 1$, $\frac{PAC_{HC}}{PAC_{SF}} - 1$, $\frac{PAC_{AD}}{PAC_{ER}} - 1$, and $\frac{PAC_{AD}}{PAC_{SF}} - 1$. Figures 5c,d,g, and h show that interactions between phases of slow frequencies (IMF5 to 8) and amplitudes of high frequencies (IMF1) are stronger in real brain networks than in simulated ER networks but weaker than in SF networks. This result is not surprising since we know that the degree distribution of brain anatomical networks do not follow a pure power law, as in SF networks, and is better described by an exponentially truncated power law [25].

The contribution of each area to the generation of the PAC phenomenon (ΔPAC) was computed following the procedure described in the previous section. Figure 6a and 6b shows the average ΔPAC across subjects for the areas with the strongest influence on PAC, for the HC and AD groups, respectively. In both groups the two areas with the strongest influence were the right superior frontal followed by the right medial orbitofrontal. We also computed the measure $1-\frac{\Delta PAC_{AD}}{\Delta PAC_{HC}}$ to determine the areas that changed more between HC and AD. Figure 6c shows the areas for which the influence on PAC was stronger in HC than in AD, whereas figure 6d displays the opposite case. We obtained that the influence of the right precentral and right superior parietal areas decreased in AD as compared to HC, whereas the influence of the right amygdala increased.

We also extracted all possible shortest paths[21] in the HC and AD brain networks, and computed the average ΔPAC of the areas involved in those paths. We found that the ΔPAC pathway *right superior frontal-right medial orbitofrontal-left superior frontal* presented the strongest ΔPAC in both HC and AD groups (red path in Fig. 7). On the other hand, the ΔPAC pathway that decreased the most during AD was *left insula-left pars opercularis-left superior temporal* (cyan in figure 7), whereas the PAC route that increased the most in AD was *right precentral-right paracentral-right precuneus* (green in figure 7). This clearly demonstrates an interhemispheric difference in PAC generation during AD.

Here, we also looked at how the scores of two customarily-used cognitive tests are related to the flow of information in AD networks as reflected by PAC. The individual clinical diagnoses assigned by the ADNI experts and used to define the HC and AD groups were based on multiple clinical evaluations [30]. The first test was the Clinical Dementia Rating Sum of Boxes (CDRSB), which provides a global rating of dementia severity through interviews on different aspects [30,31]. An algorithm conduces to a score in each of the domain boxes, which are later summed. The final score ranges from 0 to 18, with a 0-value meaning "Normal". CDRSB is a gold standard for the assessment of functional impairment [30]. The second test was the Functional Activities Questionnaire (FAQ), where an informant is asked to rate the subject's ability to perform 10 different activities of daily living [32]. The total score ranges from 0 (independent) to 30 (dependent). For each brain area the linear fit between ΔPAC and CDRSB, and ΔPAC and FAQ was computed. Figure 8 shows the linear fits in the left y-axis (colored in blue) corresponding to the regions with the strongest correlations. For the case of CDRSB, the brain areas were left middle temporal ($r = 0.61$, $p = 0.0005$), left inferior temporal ($r = 0.53$, $p = 0.004$), and right middle temporal ($r = 0.40$, $p = 0.032$), whereas for the case of FAQ, the left middle temporal ($r = 0.55$, $p = 0.002$), left inferior temporal ($r = 0.47$, $p = 0.011$) were obtained again, with the appearance of the left pars orbitalis ($r = 0.36$, $p = 0.056$) among the top-three now.

We also performed a linear fit for the two cognitive test and the strength of each area (defined as the sum of all the connections associated with area $i, s_i = \sum_{j=1}^{N} w_{ij}$). The results are displayed in the right axis (colored in red) of every panel in figure 8. We obtained the best fits for the same areas that resulted from using ΔPAC. The above-mentioned result is expected since PAC is obtained as a result of the movement of the random walkers on top of the matrices $\boldsymbol{W}$. However, the correlation values obtained were smaller and statistically significant only in two out of the six cases: the CDRSB test with the strength of left middle temporal ($r = 0.42$, $p = 0.0028$) and left inferior temporal ($r = 0.49$, $p = 0.008$) areas. The correlation between CDRSB and the right middle temporal area ($r = 0.23$, $p = 0.237$) was not significant, and neither were the correlations between the three areas and the FAQ test: left middle temporal ($r = 0.31$, $p = 0.109$), left inferior temporal ($r = 0.34, p = 0.077$), left pars orbitalis ($r = -0.16, p = 0.411$). These results suggest the existence of a relationship between cognitive impairment, functional decline and behavioral symptoms that characterize AD and the perturbations to the information flow in brain networks, as characterized by cross-frequency interactions and not by broadband interactions (functional connectivity).

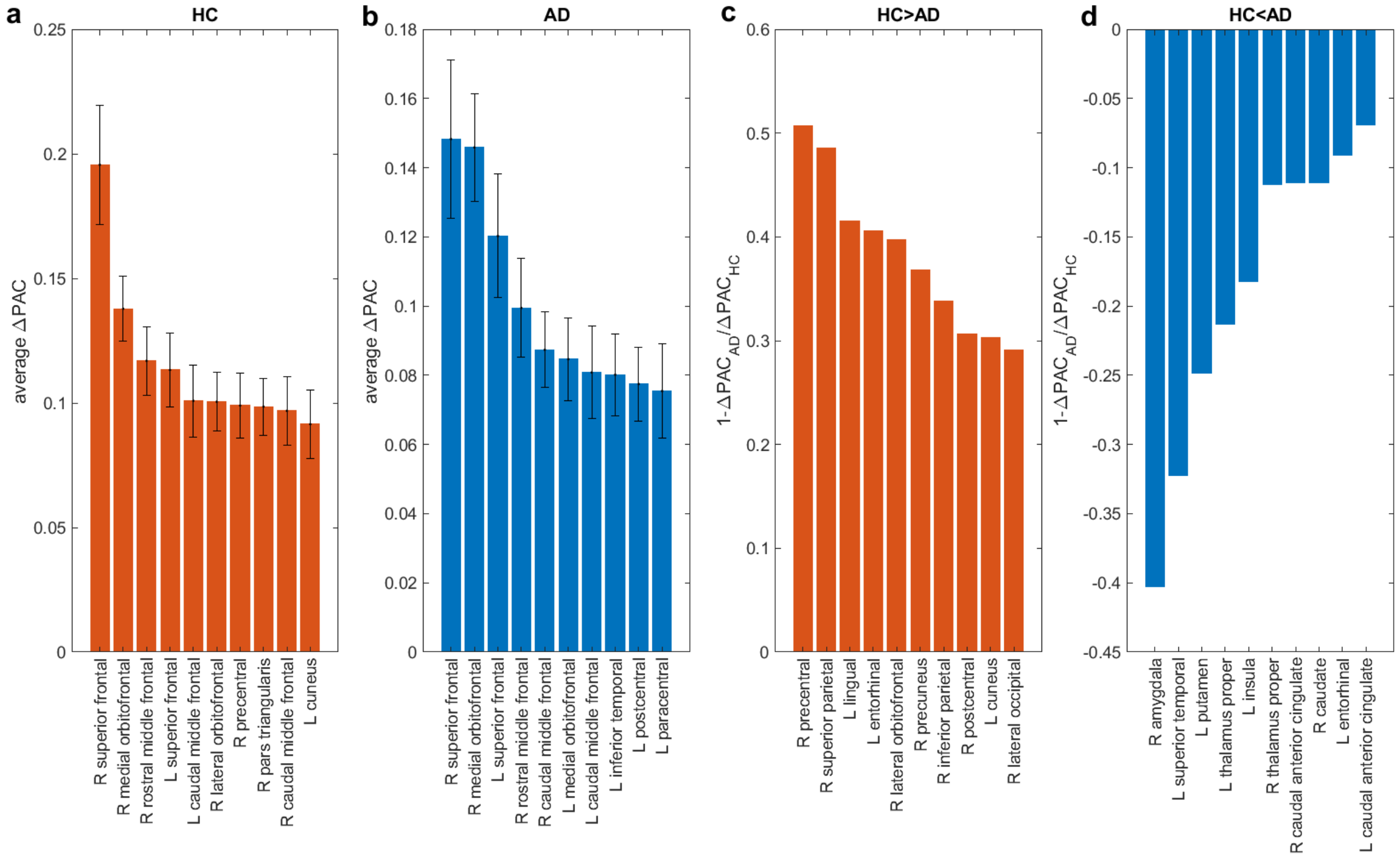

**Fig. 6.** Influence of brain areas on PAC: areas that when removed from the network change PAC the most in HC (**a**), AD (**b**). Areas for which the change was larger in HC (AD) than in AD (HC) appear in (**c**) ((**d**)). “L” and “R” denote left and right hemispheres, respectively.

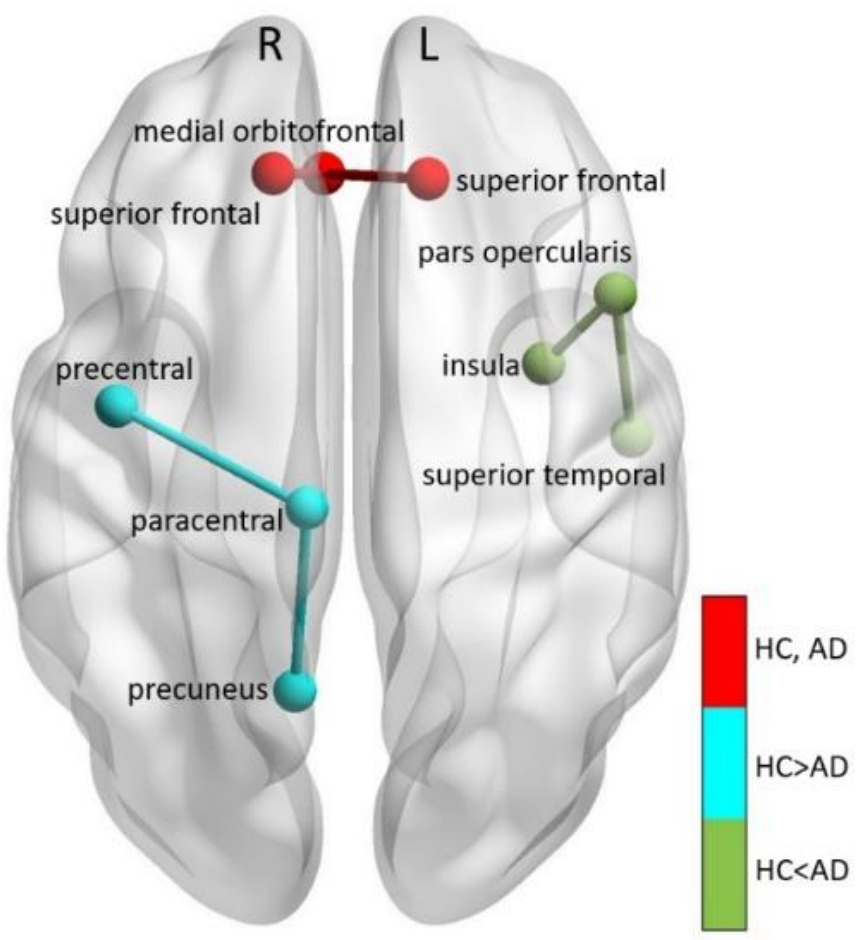

**Fig. 7.** Main PAC paths in HC and AD. Three main paths were found: 1) *right superior frontal-right medial orbitofrontal-left superior frontal* corresponding to the strongest PAC in HC, remaining also the strongest in AD (colored in red), 2) *left insula-left pars opercularis-left superior temporal*, the path that decreased the most in AD compared to HC (colored in cyan), and 3) *right precentral-right paracentral-right precuneus*, which increased the most in AD compared to HC (colored in green). “L” and “R” denote left and right hemispheres, respectively.

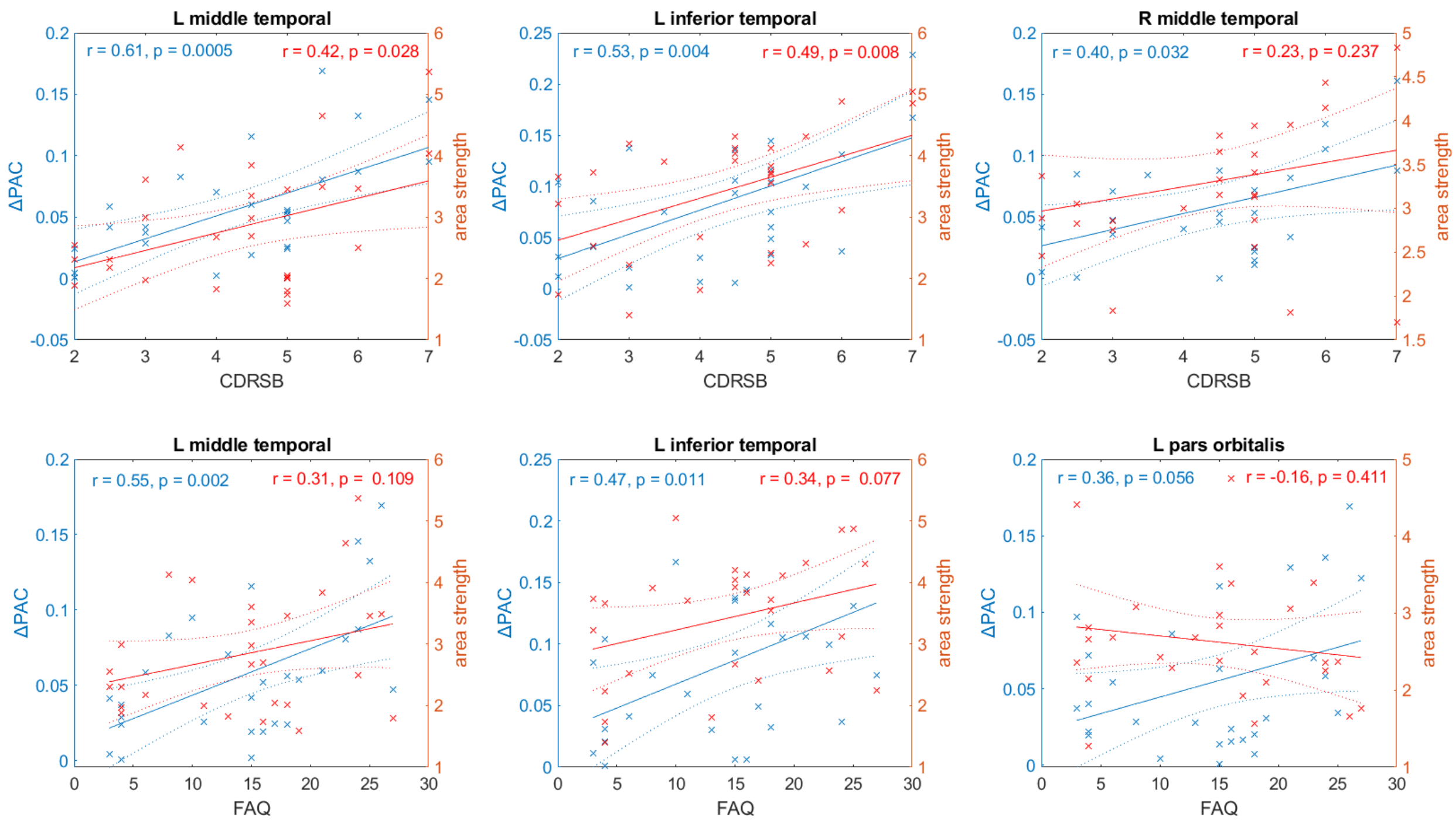

**Fig. 8.** Relationship between two cognitive tests– CDRSB and FAQ– and PAC (in blue) and values of the node strength (in red) for selected areas of the AD networks. Solid and dashed lines represent the linear fit and confidence intervals, respectively.

## Discussion

In summary, we employed random walkers to sample the spatial structure of complex networks and converted their diffusion into time series. To estimate the different temporal scales, these time series were further decomposed into intrinsic mode functions, or IMFs by means of the empirical mode decomposition technique [10]. Expressed in IMFs, the temporal scales have well-behaved Hilbert transforms [10], from which the instantaneous phases and amplitudes can be calculated. Another advantage of using EMD is that it is an adaptive and data-driven method that does not require prior knowledge on the number of temporal modes embedded into the time series. The interaction between IMFs, or cross-frequency coupling, was analysed, obtaining that cross-frequency interactions were stronger in SF than in ER networks, especially for the case of phase-amplitude coupling. SF networks presented strong PAC between slow and high frequency components of the diffusion process, whereas ER networks presented the strongest PAC between slow-frequency components. Since EMD acts essentially as a dyadic filter bank [33], some overlapping between consecutive IMFs is expected, which can result in strong CFC. This phenomenon can be seen in Supplementary Fig. 1 for the cases of PAC (interaction between the phase of IMF7 and the amplitude of IMF6), AAC (interaction between the amplitudes of IMF2 and IMF1) and PPC (interaction between the phases of IMF2 and IMF1). When filtering the data using non-overlapping bands (Supplementary Fig. 2) the strength of these couplings decreased, but the CFC patterns, specifically the strong PAC connection between slow phases (IMFs 5 to 7) and fast frequencies (IMF1) was preserved, supporting the use of EMD in our analysis.

Given a complex network, it is of interest to determine which nodes contribute the most to CFC. We studied in more detail the generation of PAC between low (IMFs 5 to 7) and high (IMF1) frequencies and found that hubs, the most connected nodes in the network were not involved in the strongest PAC (with the exception of one super-hub [23]). The most significant influence on PAC was exerted by a group of non-hubs, which connected with high probability to high degree nodes (figure 4). This facilitated the generation of PAC (information flow from low to high frequencies [13]) since low and high degree nodes were generally associated with low and high frequencies, respectively.

We applied the methodology to brain networks from HC subjects and AD patients and found that PAC activity between slow frequencies and IMF1 decreased during AD. The IMFs obtained from simulated ER and SF networks correspond to different oscillatory modes, with normalized frequencies (figure 1f and 1g). In the case of HC and AD networks, it is tempting to analyse the frequencies in Hz, in order to compare the frequency range of the different IMFs to the known frequency bands registered in the human brain. For this, we need to know the conduction delays for signals coming from different brain areas. Delays can range from a few milliseconds to several hundreds of milliseconds depending on the regions involved and the species considered [34–36]. For the human brain, there is lack of information about conduction delays between all the combinations of areas. However, if we assume the same delay of 2ms for all connections, we obtain the estimated IMFs (Supplementary Fig. 3a) corresponding roughly to frequency bands known in the EEG (delta, theta, alpha, beta, low-gamma, middle-gamma, and fast-gamma). If the delays are increased to 10ms, the frequencies of the IMFs decrease (Supplementary Fig. 3b).

When analysing the influence of specific brain areas, we found the right superior frontal and the right medial frontal to be the areas that contributed more to PAC in both HC and AD subjects. These areas belong to the default mode network (DMN), a collection of brain structures which intertwined activity increases in the absence of a task and has been associated with memory consolidation. The right superior frontal and the right medial frontal are also involved in the strongest PAC-based information flow pathway found in AD and HC: *right superior frontal-right medial orbitofrontal-left superior frontal.* The DMN is of interest to AD research given the amyloid deposits found in its regions [37,38]. We also found that the influence of the right amygdala on PAC increased during AD (figure 6d); the amygdala is known to be severely affected in AD [39]. Our results also demonstrated a marked interhemispheric difference in the generation of PAC, with areas within the left hemisphere being more correlated to the cognitive scores (figure 8). Furthermore, the PAC pathway that decreased the most during AD consisted of left hemisphere areas only (*left insula-left pars opercularis-left superior temporal*), whereas the PAC pathway that increased the most in AD was formed by areas from the right hemisphere (*right precentral-right paracentral-right precuneus*). A tentative explanation is that the brain must enhance traffic over this specific pathway we have obtained to maintain at least a minimal information flow on the right hemisphere in AD. The interhemispheric functional disconnection suggested by our results has been previously reported in mild cognitive impairment and AD subjects [40–42], and has been associated with white matter degeneration [40].

One important challenge for the AD research field is the development of efficient biomarkers. Neuroimaging biomarkers in AD are based on brain signals such as MRI, fMRI, and Positron Emission Tomography (PET). For instance, there is a consistently reported decrease in resting-state functional connectivity in AD patients compared to HCs in the DMN [43]. However, when we correlated the strength of functional connections with the reported scores of two different cognitive tests usually employed to diagnose AD, only two areas (both from the DMN), the left middle temporal and left inferior temporal presented significant correlations (0.42 and 0.49, respectively, with $p<0.05$) with one of the tests, the CDRSB. On the other hand, these two same areas presented significant and stronger correlations between PAC and both tests, the CDRSB ($r = 0.61, r = 0.53$) and FAQ ($r = 0.55, r = 0.47$). Additionally, the right middle temporal PAC presented a significant correlation ($r = 0.40$) with the CDRSB scores. These findings suggest that PAC is more sensitive to changes induced by AD than functional connectivity values and thus may be a useful biomarker for the disease.

## Materials and Methods

### Ethics statement

The study was conducted according to Good Clinical Practice guidelines, the Declaration of Helsinki Principles, US 21CFR Part 50-Protection of Human Subjects, and Part 56-Institutional Review Boards, and pursuant to state and federal HIPAA regulations (adni.loni.usc.edu). Study subjects and/or authorized representatives gave written informed consent at the time of enrollment for sample collection and completed questionnaires approved by each participating sites Institutional Review Board. The authors obtained approval from the ADNI Data Sharing and Publications Committee for data use and publication, see documents http://adni.loni.usc.edu/wpcontent/uploads/how_to_apply/ADNI_Data_Use_Agreement.pdf and http://adni.loni.usc.edu/wpcontent/uploads/how_to_apply/ADNI_Manuscript_Citations.pdf, respectively.

## Data description and processing

*Structural MRI*

Brain structural T1-weighted 3D images were acquired for all subjects. For a detailed description of acquisition details, see http://adni.loni.usc.edu/methods/documents/mriprotocols/. All images underwent non-uniformity correction using the N3 algorithm[44]. Next, they were segmented into grey matter, white matter and cerebrospinal fluid (CSF) probabilistic maps, using SPM12 (www.fil.ion.ucl.ac.uk/spm). Grey matter segmentations were standardized to MNI space [45] using the DARTEL tool [46]. Each map was modulated to preserve the total amount of signal/tissue. Mean grey matter density and determinant of the Jacobian (DJ) [46] values were calculated for 78 regions covering all the brain's grey matter [47].

*Diffusion weighted MRI*

High angular resolution diffusion imaging (HARDI) data was acquired for 51 HC subjects from ADNI. For each diffusion scan, 46 separate images were acquired, including 5 *b*0 images (no diffusion sensitization) and 41 diffusion-weighted images ($b$ = 1000 *s*/*mm*2). Other acquisition parameters were: 256 × 256 matrix, voxel size: 2.7 × 2.7 × 2.7 *mm*3, *TR* = 9000 *ms*, 52 contiguous axial slices, and scan time, 9 *min*. ADNI aligned all raw volumes to the average *b*0 image, corrected head motion and eddy current distortion.

*Anatomical networks*

Probabilistic axonal connectivity values between each brain voxel and the surface of each considered gray matter region were estimated using a fully automated fiber tractography algorithm [25] and the intravoxel fiber distributions (ODFs) of 51 HC subjects from ADNI. ODF reconstructions were based on Spherical Deconvolution[48]. A maximum of 500 *mm* trace length and a curvature threshold of ±90° were imposed as tracking parameters. Based on the resulting voxel-region connectivity maps, the individual region-region anatomical connection density matrices [25,49] were calculated. For any subject and pair of regions $i$ and $j$, the $ACD_{i,j}$ measure (0 $\leq ACD_{i,j} \leq 1$, $ACD_{i,j} \equiv ACD_{j,i}$) reflects the fraction of the region's surface involved in the axonal connection with respect to the total surface of both regions. A network backbone, containing the dominant connections in the average network, was computed using a minimum-spanning tree based algorithm [26].

*Resting fMRI acquisition/processing*

Resting-state functional images were obtained from 31 Alzheimer's disease (AD) patients and 44 HCs from ADNI using an echo-planar imaging sequence on a 3.0-Tesla Philips MRI scanner. Acquisition parameters were: 140 time points, repetition time (TR)=3000 ms, echo time (TE)=30 ms, flip angle=80°, number of slices=48, slice thickness= 3.3 mm, spatial resolution=3×3×3 mm3 and in plane matrix= 64×64. Preprocessing steps included: 1) motion correction, 2) slice timing correction, 3) spatial normalization to MNI space using the registration parameters obtained for the structural T1 image with the nearest acquisition date, and 4) signal filtering to keep only low frequency fluctuations (0.01–0.08 Hz)[50].

**Empirical mode decomposition**

EMD is a nonlinear method that decomposes a signal into its fundamental modes of oscillations, called intrinsic mode functions or IMFs. An IMF satisfies two criteria: 1) the number of zero-crossings and extrema are either equal or differ by one, and 2) the mean of its upper and lower envelopes is zero. Thus, to be successfully decomposed into IMFs, a signal must have at least one maximum and one minimum. The sifting process of decomposing a signal $x(t)$ into its IMFs is described by the following algorithm [10]:

1. All extrema are identified, and upper, $x_u(t)$, and lower, $x_l(t)$, envelopes are constructed by means of cubic spline interpolation.
2. The average of the two envelopes is subtracted from the data:

$$d(t) = x(t) - \big(x_u(t) + x_l(t)\big)/2.$$

3. The process for $d(t)$ is repeated until the resulting signal satisfies the criteria of an IMF. This first IMF is denoted as $IMF_1(t)$. The residue $r_1(t) = x(t) - IMF_1(t)$ is treated as the new data.

4. Repeat steps 1 to 3 on the residual $r_j(t)$ to obtain all the IMFs of the signal:

$$r_j(t) = x(t) - IMF_1(t) - IMF_2(t) - \cdots - IMF_j(t).$$

The procedure ends when $r_j(t)$ is a constant, a monotonic slope, or a function with only one extreme.

As a result of the EMD method, the signal $x(t)$ is decomposed into $M$ IMFs:

$$x(t) = \sum_{j=1}^{M} IMF_j(t) + r(t) \tag{1}$$

where $r(t)$ is the final residue.

A major limitation of the classical EMD method is the common presence of mode mixing, which is when one IMF consists of signals of widely disparate scales, or when a signal of a similar scale resides in different IMFs [51]. To address this issue, the ensemble empirical mode decomposition (EEMD) considers that the true IMF components are the mean of an ensemble of trials, each consisting of the signal plus a white noise of finite amplitude [51]. A more recent method, ICEEMDAN (Improved Complete Ensemble Empirical Mode Decomposition with Adaptive Noise) was built on this idea [52]. In this paper, we use the ICEEMDAN method with standard parameter values [52], which reduces the number of ensembles needed and increases the accuracy rate while avoiding spurious modes.

After computing the IMFs, the Hilbert transform can be applied to each IMF. Thus, equation (1) can be rewritten as:

$$x(t) = Real\{\sum_{j=1}^{M} a_j(t)e^{i\varphi_j(t))}\} + r(t), \quad (2)$$

where $\varphi_j(t)$ and $a_j(t)$ are the instantaneous phases and amplitudes of IMF $j$.

**Computation of cross-frequency coupling measures**

PAC is the phenomenon where the instantaneous phase of a low frequency oscillation modulates the instantaneous amplitude of a higher frequency oscillation [14] [15]. To compute PAC, we used the modification to the mean-vector length modulation index [53]:

$$PAC = \left|\frac{1}{N}\sum_{n=1}^{N} a_2(n)\left(e^{i\varphi_1(n)} - \overline{\varphi}\right)\right|, \ \overline{\varphi} = \frac{1}{N}\sum_{n=1}^{N} e^{i\varphi_1(n)} \quad (3)$$

where $N$ is the total number of time points, $a_2$ is the amplitude of the modulated signal, $\varphi_1$ is the phase of the modulating signal, and $\overline{\varphi}$ is a factor introduced to remove phase clustering bias.

PPC, which corresponds to the synchronization between two instantaneous phases[17], was calculated by using the *n:m* phase-locking value(PLV) [54]:

$$PPC = \left|\frac{1}{T}\sum_{t=1}^{T} e^{i\left(n\varphi_1(t) - m\varphi_2(t)\right)}\right| \quad (4)$$

where $\varphi_1$ and $\varphi_2$ are the instantaneous phases, and *m* and *n* are integers. We tested all possible combinations of $n$ and $m$ for $n = 1,2, \ldots, 30$, $m = 1,2, \ldots, 30$, with $m > n$, and selected the one producing the highest PPC value.

AAC, the co-modulation of the instantaneous amplitudes $a_1$ and $a_2$ of two signals, was estimated by means of their the correlation [16]:

$$AAC = corr(a_1(n), a_2(n)) \quad (5)$$

A significance value can be attached to any of the above measures through a surrogate data approach where we offset $\varphi_1$ and $a_1$ by a random time lag. We can thus compute 1000 surrogate PAC, PPC, and AAC values. From the surrogate dataset, we first computed the mean $\mu$ and standard deviation $\sigma$, and then computed a Z-score as:

$$Z_{PAC} = \frac{PAC - \mu_{PAC}}{\sigma_{PAC}}, Z_{PPC} = \frac{PPC - \mu_{PPC}}{\sigma_{PPC}}, \; Z_{AAC} = \frac{AAC - \mu_{AAC}}{\sigma_{AAC}} \quad (6)$$

The normal distribution of the surrogated data was tested with the Jarque-Bera test, and the *p*-value that corresponded to the standard Gaussian variate was also computed. P-values were corrected by means of a multiple comparison analysis based on the false discovery rate (FDR) [55].

## Acknowledgments

This work was partially supported by grant RGPIN-2015-05966 from the Natural Sciences and Engineering Research Council of Canada. Data collection and sharing for this project was funded by the Alzheimer's Disease Neuroimaging Initiative (ADNI) (National Institutes of Health Grant U01 AG024904) and DOD ADNI (Department of Defense award number W81XWH-12-2-0012). ADNI is funded by the National Institute on Aging, the National Institute of Biomedical Imaging and Bioengineering, and through generous contributions from the following: AbbVie, Alzheimer's Association; Alzheimer's Drug Discovery Foundation; Araclon Biotech; BioClinica, Inc.; Biogen; Bristol-Myers Squibb Company; CereSpir, Inc.; Eisai Inc.; Elan Pharmaceuticals, Inc.; Eli Lilly and Company; EuroImmun; F. Hoffmann-La Roche Ltd and its affiliated company Genentech, Inc.; Fujirebio; GE Healthcare; IXICO Ltd.; Janssen Alzheimer Immunotherapy Research & Development, LLC.; Johnson & Johnson Pharmaceutical Research & Development LLC.; Lumosity; Lundbeck; Merck and Co., Inc.; Meso Scale Diagnostics, LLC.; NeuroRx Research; Neurotrack Technologies; Novartis Pharmaceuticals Corporation; Pfizer Inc.; Piramal Imaging;

Servier; Takeda Pharmaceutical Company; and Transition Therapeutics. The Canadian Institutes of Health Research is providing funds to support ADNI clinical sites in Canada. Private sector contributions are facilitated by the Foundation for the National Institutes of Health (www.fnih.org). The grantee organization is the Northern California Institute for Research and Education, and the study is coordinated by the Alzheimer's Disease Cooperative Study at the University of California, San Diego. ADNI data are disseminated by the Laboratory for Neuro Imaging at the University of Southern California. Data used in preparation of this article were obtained from the ADNI database (adni.loni.usc.edu). As such, the investigators within the ADNI contributed to the design and implementation of ADNI and/or provided data but did not participate in analysis or writing of this report.

**Data availability**. All MRI and fMRI data used in this study were obtained from the Alzheimer's Disease Neuroimaging Initiative (ADNI) database (http://adni.loni.usc.edu). For researchers who meet the criteria for access to the data; access to the ADNI data is available through an online application, which can be submitted at the following link: http://adni.loni.usc.edu/data-samples/access-data/.

**Author Contributions**

RCS conceived the project, analyzed and interpreted the results, and drafted the manuscript.

LSR, MD, YIM and JSB assisted with analysis and interpretation of data. All authors assisted with editing of the manuscript.

## Additional Information

The authors declare they have no competing interests or other interests that might be perceived to influence the interpretation of the article.

## References

1. Gallos, L. K., Song, C., Havlin, S. & Makse, H. A. Scaling theory of transport in complex biological networks. *Proc. Natl. Acad. Sci. U. S. A.* **104,** 7746–51 (2007).
2. Raj, A., Kuceyeski, A. & Weiner, M. A Network Diffusion Model of Disease Progression in Dementia. *Neuron* **73,** 1204–1215 (2012).
3. Gfeller, D., De Los Rios, P., Caflisch, A. & Rao, F. Complex network analysis of free-energy landscapes. *Proc. Natl. Acad. Sci.* **104,** 1817–1822 (2007).
4. Simonsen, I., Astrup Eriksen, K., Maslov, S. & Sneppen, K. Diffusion on complex networks: a way to probe their large-scale topological structures. *Phys. A Stat. Mech. its Appl.* **336,** 163–173 (2004).
5. Pearson, K. The Problem of the Random Walk. *Nature* **72,** 294–294 (1905).
6. Noh, J. D. & Rieger, H. Random Walks on Complex Networks. *Phys. Rev. Lett.* **92,** 118701 (2004).
7. Noskowicz, S. H. & Goldhirsch, I. First-passage-time distribution in a random random walk. *Phys. Rev. A* **42,** 2047–2064 (1990).
8. Tejedor, V., Bénichou, O. & Voituriez, R. Global mean first-passage times of random walks on complex networks. *Phys. Rev. E* **80,** 065104 (2009).
9. Bonaventura, M., Nicosia, V. & Latora, V. Characteristic times of biased random walks on complex networks. *Phys. Rev. E* **89,** 012803 (2014).
10. Huang, N. E. *et al.* The empirical mode decomposition and the Hilbert spectrum for nonlinear and non-stationary time series analysis. *Proc. R. Soc. A Math. Phys. Eng. Sci.* **454,** 903–995 (1998).
11. Erdös & Rényi, A. On random graphs, I. *Publ. Math.* **6,** (1959).
12. Barabasi, A.-L. & Albert, R. Emergence of scaling in random networks. *Science* **286,** 509–12 (1999).
13. Sotero, R. C. Topology, Cross-Frequency, and Same-Frequency Band Interactions Shape the Generation of Phase-Amplitude Coupling in a Neural Mass Model of a Cortical Column. *PLOS Comput. Biol.* **12,** e1005180 (2016).
14. Sotero, R. C. *et al.* Laminar Distribution of Phase-Amplitude Coupling of Spontaneous Current Sources and Sinks. *Front. Neurosci.* **9,** 454 (2015).
15. Sotero, R. C. Modeling the Generation of Phase-Amplitude Coupling in Cortical Circuits: From Detailed Networks to Neural Mass Models. *Biomed Res. Int.* **2015,** 1–12 (2015).
16. Bruns, A. & Eckhorn, R. Task-related coupling from high- to low-frequency signals among visual cortical areas in human subdural recordings. *Int. J. Psychophysiol.* **51,** 97–116 (2004).
17. Lachaux, J.-P., Rodriguez, E., Martinerie, J. & Varela, F. J. Measuring phase synchrony in brain signals. *Hum. Brain Mapp.* **8,** 194–208 (1999).

18. Barrat, A., Barthelemy, M. & Vespignani, A. *Dynamical Processes on Complex Networks*. (Cambridge University Press, 2008). doi:10.1017/CBO9780511791383
19. Taylor, A. & Higham, D. J. CONTEST. *ACM Trans. Math. Softw.* **35,** 1–17 (2009).
20. Colominas, M. A., Schlotthauer, G. & Torres, M. E. Improved complete ensemble EMD: A suitable tool for biomedical signal processing. *Biomed. Signal Process. Control* **14,** 19–29 (2014).
21. Rubinov, M. & Sporns, O. Complex network measures of brain connectivity: Uses and interpretations. *Neuroimage* **52,** 1059–1069 (2010).
22. Sporns, O., Honey, C. J. & Kötter, R. Identification and Classification of Hubs in Brain Networks. *PLoS One* **2,** e1049 (2007).
23. Hao, D., Ren, C. & Li, C. *Revisiting the variation of clustering coefficient of biological networks suggests new modular structure*. (2012). doi:10.1186/1752-0509-6-34
24. Sanchez-rodriguez, L. M. *et al.* Design of optimal nonlinear network controllers for Alzheimer ' s disease. 1–24 (2018). doi:10.1371/journal.pcbi.1006136
25. Iturria-Medina, Y., Sotero, R. C., Canales-Rodriguez, E. J., Aleman-Gomez, Y. & Melie-Garcia, L. Studying the human brain anatomical network via diffusion-weighted MRI and Graph Theory. *Neuroimage* **40,** 1064–1076 (2008).
26. Hagmann, P. *et al.* Mapping the Structural Core of Human Cerebral Cortex. *PLoS Biol.* **6,** e159 (2008).
27. Cao, J., Wang, X., Liu, H. & Alexandrakis, G. Directional changes in information flow between human brain cortical regions after application of anodal transcranial direct current stimulation (tDCS) over Broca's area. *Biomed. Opt. Express* **9,** 5296–5317 (2018).
28. Yanagawa, T., Chao, Z. C., Hasegawa, N. & Fujii, N. Large-Scale Information Flow in Conscious and Unconscious States: an ECoG Study in Monkeys. *PLoS One* **8,** e80845 (2013).
29. Zhang, Z., Shan, T. & Chen, G. Random walks on weighted networks. *Phys. Rev. E* **87,** 012112 (2013).
30. Defina, P. A., Moser, R. S., Glenn, M., Lichtenstein, J. D. & Fellus, J. Alzheimer's disease clinical and research update for health care practitioners. *J. Aging Res.* **2013,** 207178 (2013).
31. Doody, R. S. *et al.* Predicting progression of Alzheimer's disease. *Alzheimers. Res. Ther.* **2,** 2 (2010).
32. Juva, K. *et al.* Functional assessment scales in detecting dementia. *Age Ageing* **26,** 393–400 (1997).
33. Flandrin, P., Rilling, G. & Goncalves, P. Empirical Mode Decomposition as a Filter Bank. *IEEE Signal Process. Lett.* **11,** 112–114 (2004).
34. Budd, J. M. L. & Kisvárday, Z. F. Communication and wiring in the cortical connectome. *Front. Neuroanat.* **6,** 42 (2012).
35. Stoelzel, C. R., Bereshpolova, Y., Alonso, J.-M. & Swadlow, H. A. Axonal Conduction Delays, Brain State, and Corticogeniculate Communication. *J. Neurosci.* **37,** 6342–6358 (2017).
36. Aboitiz, F., López, J. & Montiel, J. Long distance communication in the human brain: timing constraints for inter-hemispheric synchrony and the origin of brain lateralization. *Biol. Res.* **36,** 89–99 (2003).
37. Sheline, Y. I. *et al.* Amyloid plaques disrupt resting state default mode network connectivity in cognitively normal elderly. *Biol. Psychiatry* **67,** 584–7 (2010).

38. Sperling, R. A. *et al.* Amyloid deposition is associated with impaired default network function in older persons without dementia. *Neuron* **63,** 178–88 (2009).
39. Poulin, S. P., Dautoff, R., Morris, J. C., Feldman Barrett, L. & Dickerson, B. C. Amygdala atrophy is prominent in early Alzheimer's disease and relates to symptom severity on behalf of the Alzheimer's Disease Neuroimaging Initiative. *Psychiatry Res* **194,** 7–13 (2011).
40. Wang, Z. *et al.* Interhemispheric Functional and Structural Disconnection in Alzheimer's Disease: A Combined Resting-State fMRI and DTI Study. *PLoS One* **10,** e0126310 (2015).
41. Qiu, Y. *et al.* Inter-hemispheric functional dysconnectivity mediates the association of corpus callosum degeneration with memory impairment in AD and amnestic MCI. *Sci. Rep.* **6,** 32573 (2016).
42. Korolev, I., Bozoki, A., Majumdar, S., Berger, K. & Zhu, D. Alzheimer's disease reduces inter-hemispheric hippocampal functional connectivity. *Alzheimer's Dement.* **7,** S739 (2011).
43. Dennis, E. L. & Thompson, P. M. Functional Brain Connectivity Using fMRI in Aging and Alzheimer's Disease. *Neuropsychol. Rev.* **24,** 49–62 (2014).
44. Sled, J. G., Zijdenbos, A. P. & Evans, A. C. A nonparametric method for automatic correction of intensity nonuniformity in MRI data. *IEEE Trans. Med. Imaging* **17,** 87–97 (1998).
45. Evans, A. C., Kamber, M., Collins, D. L. & MacDonald, D. in *Magnetic Resonance Scanning and Epilepsy* 263–274 (Springer US, 1994). doi:10.1007/978-1-4615-2546-2_48
46. Ashburner, J. A fast diffeomorphic image registration algorithm. *Neuroimage* **38,** 95–113 (2007).
47. Klein, A. & Tourville, J. 101 Labeled Brain Images and a Consistent Human Cortical Labeling Protocol. *Front. Neurosci.* **6,** 171 (2012).
48. Tournier, J.-D. *et al.* Resolving crossing fibres using constrained spherical deconvolution: Validation using diffusion-weighted imaging phantom data. *Neuroimage* **42,** 617–625 (2008).
49. Sotero, R. C., Trujillo-Barreto, N. J., Iturria-Medina, Y., Carbonell, F. & Jimenez, J. C. Realistically Coupled Neural Mass Models Can Generate EEG Rhythms. *Neural Comput.* **19,** 478–512 (2007).
50. Yan, C. & Zang, Y. DPARSF: a MATLAB toolbox for "pipeline" data analysis of resting-state fMRI. *Front. Syst. Neurosci.* **4,** 13 (2010).
51. WU, Z. & HUANG, N. E. ENSEMBLE EMPIRICAL MODE DECOMPOSITION: A NOISE-ASSISTED DATA ANALYSIS METHOD. *Adv. Adapt. Data Anal.* **01,** 1–41 (2009).
52. Colominas, M. A., Schlotthauer, G. & Torres, M. E. Improved complete ensemble EMD: A suitable tool for biomedical signal processing. *Biomed. Signal Process. Control* **14,** 19–29 (2014).
53. van Driel, J., Cox, R. & Cohen, M. X. Phase-clustering bias in phase–amplitude cross-frequency coupling and its removal. *J. Neurosci. Methods* **254,** 60–72 (2015).
54. Tass, P. *et al.* Detection of n : m Phase Locking from Noisy Data: Application to Magnetoencephalography. *Phys. Rev. Lett.* **81,** 3291–3294 (1998).
55. Benjamini, Y. & Hochberg, Y. *Controlling the False Discovery Rate: A Practical and Powerful Approach to Multiple Testing. Source: Journal of the Royal Statistical Society.*

*Series B (Methodological)* **57,** (1995).

## Consortia

Alzheimer's Disease Neuroimaging Initiative (ADNI):

Michael W. Weiner[5], Paul Aisen[6], Ronald Petersen[6], Clifford R. Jack[6], William Jagust[8], John Q. Trojanowki[9],Arthur W. Toga[10], Laurel Beckett[11], Robert C. Green[12], Andrew J. Saykin[13], John Morris[14], Leslie M. Shaw[10],Zaven Khachaturian[6,15], Greg Sorensen[16], Lew Kuller[17], Marc Raichle[14], Steven Paul[18], Peter Davies[19],Howard Fillit[20], Franz Hefti[21], Davie Holtzman[14], M. Marcel Mesulam[22], William Potter[23], Peter Snyder[24],Adam Schwartz[25], Tom Montine[26], Ronald G. Thomas[6], Michael Donohue[6], Sarah Walter[6], Devon Gessert[6],Tamie Sather[6], Gus Jiminez[6], Danielle Harvey[11], Matthew Bernstein[7], Nick Fox[27], Paul Thompson[28],Norbert Schuff[5,11], Bret Borowski[7], Jeff Gunter[7], Matt Senjem[7], Prashanthi Vemuri[7], David Jones[7],Kejal Kantarci[7], Chad Ward[7], Robert A. Koeppe[29], Norm Foster[30], Eric M. Reiman[31], Kewei Chen[31], Chet Mathis[17], Susan Landau[8], Nigel J. Cairns[14], Erin Householder[14], Lisa Taylor-Reinwald[14], Virginia Lee[9],Magdalena Korecka[9], Michal Figurski[9], Karen Crawford[10], Scott Neu[10], Tatiana M. Foroud[13], Steven Potkin[32],Li Shen[13], Kelley Faber[13], Sungeun Kim[13], Kwangsik Nho[13], Leon Thal[6], Neil Buckholtz[33], Marylyn Albert[34],Richard Frank[35], John Hsiao[33], Jeffrey Kaye[36], Joseph Quinn[36], Betty Lind[36], Raina Carter[36], Sara Dolen[36],Lon S. Schneider[10], Sonia Pawluczyk[10], Mauricio Beccera[10], Liberty Teodoro[10], Bryan M. Spann[10], James Brewer[6],Helen Vanderswag[6], Adam Fleisher[6],[31], Judith L. Heidebrink[29], Joanne L. Lord[29], Sara S. Mason[7],Colleen S. Albers[7], David Knopman[7], Kris Johnson[7], Rachelle S. Doody[37], Javier Villanueva-Meyer[37],Munir Chowdhury[37], Susan Rountree[37], Mimi Dang[37], Yaakov Stern[37], Lawrence S. Honig[37], Karen L. Bell[37],Beau Ances[14], Maria Carroll[14], Sue Leon[14], Mark A. Mintun[14], Stacy Schneider[14], Angela Oliver[14],Daniel Marson[38], Randall Griffith[38], David Clark[38], David Geldmacher[38], John Brockington[38], Erik Roberson[38],Hillel Grossman[39], Effie Mitsis[39], Leyla de Toledo-Morrell[40], Raj C. Shah[40], Ranjan Duara[41], Daniel Varon[41],Maria T. Greig[41], Peggy Roberts[41], Marilyn Albert[34], Chiadi Onyike[34], Daniel D'Agostino[34], Stephanie Kielb[34],James E. Galvin[42], Brittany Cerbone[42], Christina A. Michel[42], Henry Rusinek[42], Mony J. de Leon[42], Lidia Glodzik[42],Susan De Santi[42], P. Murali Doraiswamy[43], Jeffrey R. Petrella[43], Terence Z.Wong[43], Steven E. Arnold[9],Jason H. Karlawish[9], David Wolk[9], Charles D. Smith[44], Greg Jicha[44], Peter Hardy[44], Partha Sinha[44],Elizabeth Oates[44], Gary Conrad[44], Oscar L. Lopez[17], MaryAnn Oakley[17], Donna M. Simpson[34],Anton P. Porsteinsson[45], Bonnie S. Goldstein[45], Kim Martin[45], Kelly M. Makino[45], M. Saleem Ismail[45],Connie Brand[45], Ruth A. Mulnard[32], Gaby Thai[32], Catherine Mc-Adams-Ortiz[32], Kyle Womack[46],Dana Mathews[46], Mary Quiceno[46], Ramon Diaz-Arrastia[46], Richard King[46], Myron Weiner[46],Kristen Martin-Cook[46], Michael DeVous[46], Allan I. Levey[47], James J. Lah[47], Janet S. Cellar[47], Jeffrey M. Burns[48],Heather S. Anderson[48], Russell H. Swerdlow[48], Liana Apostolova[28], Kathleen Tingus[28], Ellen Woo[28],Daniel H.S. Silverman[28], Po H. Lu[28], George Bartzokis[28], Neill R. Graff-Radford[49], Francine Parfitt[49],Tracy Kendall[49], Heather Johnson[49], Martin R. Farlow[13], AnnMarie Hake[13], Brandy R. Matthews[13],Scott Herring[13], Cynthia Hunt[13], Christopher H. van Dyck[50], Richard E. Carson[50], Martha G. MacAvoy[50],Howard Chertkow[51], Howard Bergman[51], Chris Hosein[51], Sandra Black[52], Bojana Stefanovic[52], Curtis Caldwell[52],Ging-Yuek Robin Hsiung[53], Howard Feldman[53], Benita Mudge[53], Michele Assaly[53], Andrew Kertesz[54,55,56],John Rogers[54,56], Charles Bernick[54], Donna Munic[54], Diana Kerwin[22], Marek-Marsel Mesulam[22],Kristine Lipowski[22], Chuang-Kuo Wu[22], Nancy Johnson[22], Carl Sadowsky[57], Walter Martinez[57], Teresa Villena[57],Raymond Scott Turner[58], Kathleen Johnson[58], Brigid Reynolds[58], Reisa A. Sperling[12], Keith A. Johnson[12],Gad Marshall[12], Meghan Frey[12], Barton Lane[12], Allyson Rosen[12], Jared Tinklenberg[12], Marwan N. Sabbagh[59],Christine M. Belden[59], Sandra A. Jacobson[59], Sherye A. Sirrel[59], Neil Kowall[60], Ronald Killiany[60],Andrew E. Budson[60], Alexander Norbash[60], Patricia Lynn Johnson[60], Joanne Allard[61], Alan Lerner[62],Paula Ogrocki[62], Leon Hudson[62], Evan Fletcher[11], Owen Carmichael[11], John Olichney[11], Charles DeCarli[11],Smita Kittur[63], Michael Borrie[64], T.-Y. Lee[64], Rob Bartha[64], Sterling Johnson[65], Sanjay

Asthana[65],Cynthia M. Carlsson[65], Steven G. Potkin[29], Adrian Preda[29], Dana Nguyen[29], Pierre Tariot[31], Stephanie Reeder[31],Vernice Bates[66], Horacio Capote[66], Michelle Rainka[66], Douglas W. Scharre[67], Maria Kataki[67], Anahita Adeli[67],Earl A. Zimmerman[68], Dzintra Celmins[68], Alice D. Brown[68], Godfrey D. Pearlson[69], Karen Blank[69],Karen Anderson[69], Robert B. Santulli[70], Tamar J. Kitzmiller[70], Eben S. Schwartz[70], Kaycee M. Sink[71],Jeff D.Williamson[71], Pradeep Garg[71], FranklinWatkins[71], Brian R. Ott[72], Henry Querfurth[72], Geoffrey Tremont[72],Stephen Salloway[73], Paul Malloy[73], Stephen Correia[73], Howard J. Rosen[5], Bruce L. Miller[5], Jacobo Mintzer[74],Kenneth Spicer[74], David Bachman[74], Elizabether Finger[56], Stephen Pasternak[56], Irina Rachinsky[56], Dick Drost[56],Nunzio Pomara[75], Raymundo Hernando[75], Antero Sarrael[75], Susan K. Schultz[76], Laura L. Boles Ponto[76],Hyungsub Shim[76], Karen Elizabeth Smith[76], Norman Relkin[18], Gloria Chaing[18], Lisa Raudin[15,18], Amanda Smith[77],Kristin Fargher[77], Balebail Ashok Raj[77], Thomas Neylan[5], Jordan Grafman[22], Melissa Davis[6],Rosemary Morrison[6], Jacqueline Hayes[5], Shannon Finley[5], Karl Friedl[78], Debra Fleischman[40],Konstantinos Arfanakis[40], Olga James[43], Dino Massoglia[74], J. Jay Fruehling[65], Sandra Harding[65],Elaine R. Peskind[26], Eric C. Petrie[67], Gail Li[67], Jerome A. Yesavage[79], Joy L. Taylor[79] and Ansgar J. Furst[79].

[5] UC San Francisco, California, USA. [6] UC San Diego, California, USA. [7] Mayo Clinic, Rochester, New York, USA. [8] UC Berkeley, California, USA. [9] UPenn, Philadelphia, Pennsylvania, USA. [10] USC, Los Angeles, California, USA. [11] UC Davis, California, USA. [12] Brigham and Women's Hospital/Harvard MedicalSchool, Boston, Massachusetts, USA. [13] Indiana University, Bloomington, Indiana, USA. [14]Washington University St Louis, Missouri, USA. [15] Prevent Alzheimer's Disease 2131, Rockville, Maryland, USA. [16] Siemens, Munich, Germany. [17] University of Pittsburg, Pennsylvania, USA. [18] Cornell University, Ithaca, New York, USA. [19] Albert Einstein College of Medicine of Yeshiva University, Bronx, New York, USA. [20]AD Drug Discovery Foundation, New York City, New York, USA. [21] Acumen Pharmaceuticals, Livermore, California, USA. [22] Northwestern University, Evanston and Chicago, Illinois, USA. [23] NationalInstitute of Mental Health, Rockville, Maryland, USA. [24] Brown University, Providence, Rhode Island, USA. [25] Eli Lilly, Indianapolis, Indiana, USA. [26] University of Washington, Seattle, Washington, USA. [27] University of London, London, England. [28] UCLA, Los Angeles, California, USA. [29] University of Michigan, AnnArbor, Michigan, USA. [30] University of Utah, Salt Lake, Utah, USA. [31] Banner Alzheimer's Institute, Phoenix, Arizona, USA. [32] UC Irvine, Irvine, California, USA.[33] National Institute on Aging, Bethesda, Maryland, USA. [34] Johns Hopkins University, Baltimore, Maryland, USA. [35] Richard Frank Consulting, Washington,DC, USA. [36] Oregon Health and Science University, Portland, Oregon, USA. [37] Baylor College of Medicine, Houston, Texas, USA. [38] University of Alabama,Birmingham, Alabama, USA. [39] Mount Sinai School of Medicine, New York City, New York, USA. [40] Rush University Medical Center, Chicago, Illinois, USA.[41] Wien Center, Miami, Florida, USA. [42] New York University, New York City, New York, USA. [43] Duke University Medical Center, Durham, North Carolina, USA. [44] University of Kentucky, Lexington, Kentucky, USA. [45] University of Rochester Medical Center, Rochester, New York, USA. [46] University of TexasSouthwestern Medical School, Dallas, Texas, USA. [47] Emory University, Atlanta, Georgia, USA. [48] University of Kansas, Medical Center, Kansas City, Kansas, USA. [49] Mayo Clinic, Jacksonville, Florida, USA. [50] Yale University School of Medicine, New Haven, Connecticut, USA. [51] McGill University/Montreal-JewishGeneral Hospital, Montreal, Quebec, Canada. [52] Sunnybrook Health Sciences, Toronto, Ontario, Canada. [53] U.B.C. Clinic for AD & Related Disorders,Vancouver, British Columbia, Canada. [54] Cognitive Neurology–St Joseph's, London, Ontario, Canada. [55] Cleveland Clinic Lou Ruvo Center for Brain Health, LasVegas, Nevada, USA. [56] St Joseph's Health Care, London, Ontario, Canada. [57] Premiere Research Institute, Palm Beach Neurology, Miami, Florida, USA.[58] Georgetown University Medical Center, Washington, DC, USA. [59] Banner Sun Health Research Institute, Sun City, Arizona, USA. [60] Boston University,Boston, Massachusetts, USA. [61] Howard University, Washington, DC, USA. [62] Case Western Reserve University, Cleveland, Ohio, USA. [63] Neurological Care of CNY, Liverpool, New York, USA. [64]

Parkwood Hospital, London, Ontario, USA. [65] University of Wisconsin, Madison, Wisconsin, USA. [66] Dent NeurologicInstitute, Amherst, New York, USA. [67] Ohio State University, Columbus, Ohio, USA. [68] Albany Medical College, Albany, New York, USA. [69] Hartford Hospital,Olin Neuropsychiatry Research Center, Hartford, Connecticut, USA. [70] Dartmouth-Hitchcock Medical Center, Lebanon, New Hampshire, USA. [71]WakeForest University Health Sciences, Winston-Salem, North Carolina, USA. [72] Rhode Island Hospital, Providence, Rhode Island, USA. [73] Butler Hospital,Providence, Rhode Island, USA. [74] Medical University South Carolina, Charleston, South Carolina, USA. [75] Nathan Kline Institute, Orangeburg, New York,USA. [76] University of Iowa College of Medicine, Iowa City, Iowa, USA. [77] University of South Florida: USF Health Byrd Alzheimer's Institute, Tampa, Florida,USA. [78] Department of Defense, Arlington, Virginia, USA. [79] Stanford University, Stanford, California, USA

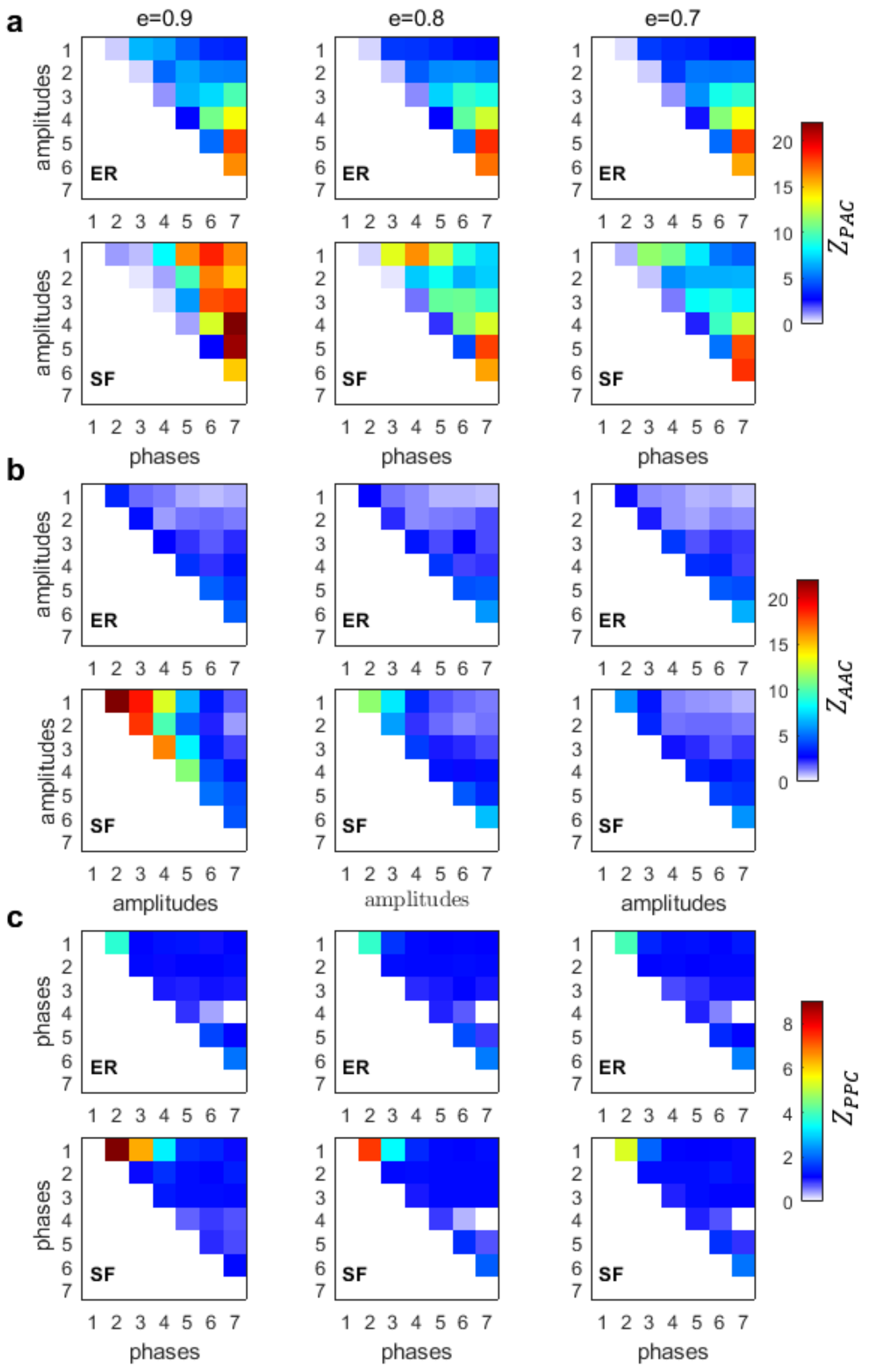

**Supplementary Fig. 1.** Z-scores for the cross-frequency interactions between the fundamental modes of diffusion (represented by IMFs) on ER and SF networks. All simulated networks had $N = 500$ nodes, and $10^4$ random walkers were placed over it, each performing $5000$ time steps. Three different values of network sparsity were considered: $e = [0.9, 0.8, 0.7]$. (**a**) phase-amplitude coupling (PAC), (**b**) amplitude-amplitude coupling (AAC), (**c**) phase-phase coupling (PPC). Non-significant $CFC$ values are displayed in white.

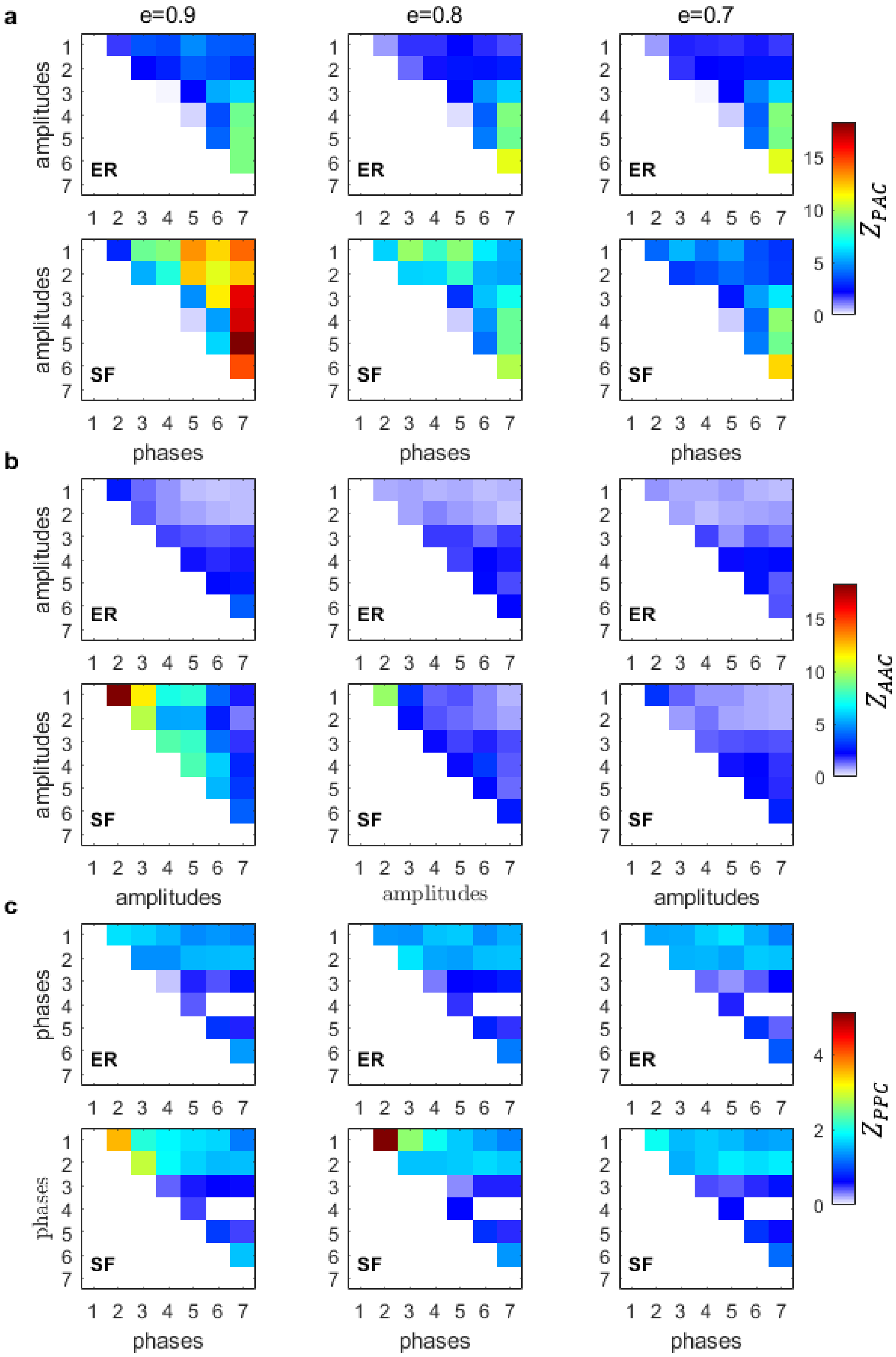

**Supplementary Fig. 2.** Z-scores for the cross-frequency interactions on ER and SF networks between non-overlapping frequency bands (0.001-0.009, 0.010-0.020, 0.021-0.040, 0.041-0.060, 0.061-0.100, 0.101-0.250, and 0.251-0.490 cycles/sample). All simulated networks had $N = 500$ nodes, and $10^4$ random walkers were placed over it, each performing $5000$ time steps. Three different values of network sparsity were considered: $e = [0.9, 0.8, 0.7]$]. (**a**) phase-amplitude coupling (PAC), (**b**) amplitude-amplitude coupling (AAC), (**c**) phase-phase coupling (PPC). Non-significant $CFC$ values are displayed in white.

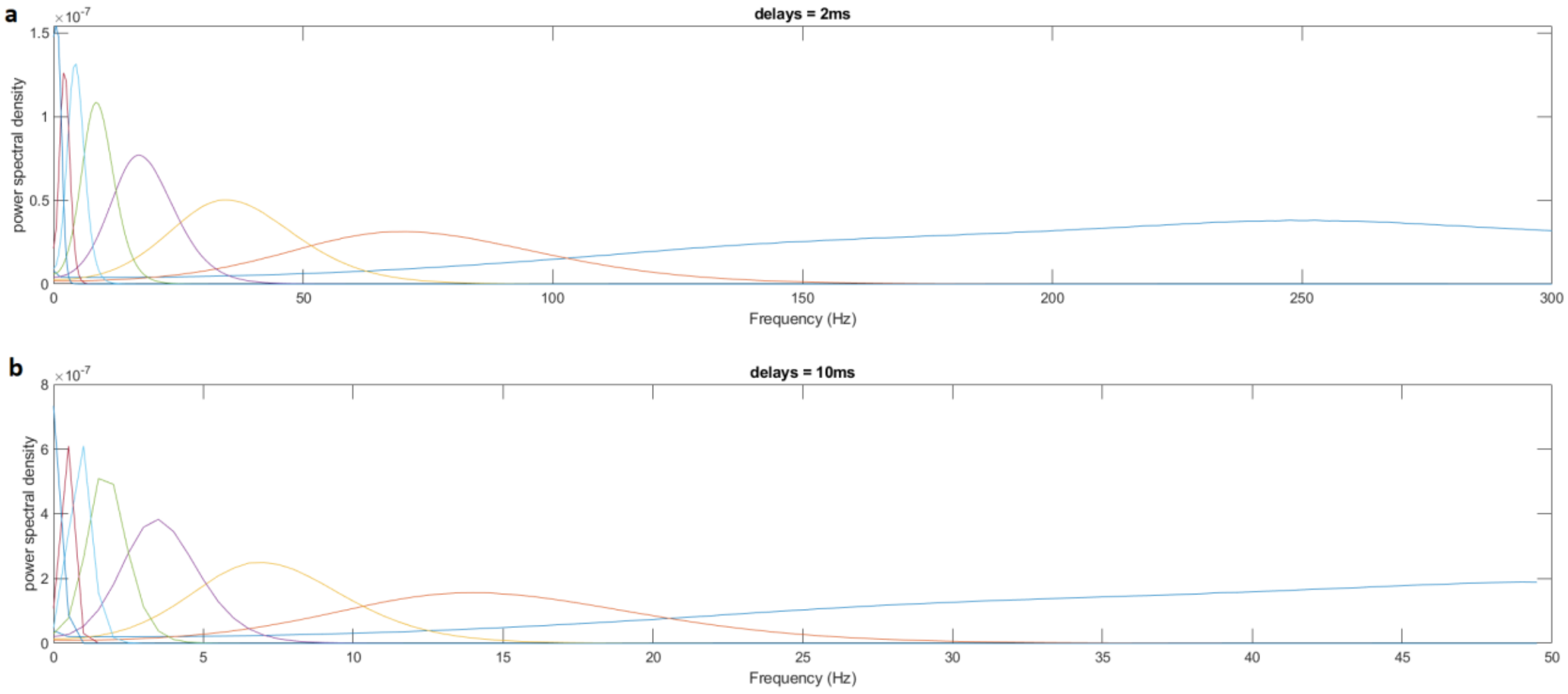

**Supplementary Fig. 3.** Average power spectral density of IMFs estimated from random walkers movement on HC networks assuming connection delays of: (**a**) 2ms, (**b**) 10ms.